\documentclass[fleqn,usenatbib]{mnras}

\usepackage{newtxtext,newtxmath}

\usepackage[T1]{fontenc}
\usepackage{ae,aecompl}

\usepackage{graphicx}	
\usepackage{amsmath}	
\usepackage{amssymb}	
\usepackage{hyperref}
\usepackage{caption}
\usepackage{subcaption}
\usepackage{tablefootnote}

\title[Environment from cross-correlations]{Environment from cross-correlations: connecting hot gas and the quenching of galaxies}

\author[E. Kukstas et al.]{
Egidijus Kukstas\thanks{E-mail: E.Kukstas@2016.ljmu.ac.uk},
Ian G. McCarthy\thanks{E-mail: i.g.mccarthy@ljmu.ac.uk},~
Ivan K. Baldry,~
Andreea S. Font
\\
Astrophysics Research Institute, Liverpool John Moores University, 146 Brownlow Hill, Liverpool L3 5RF\\
}

\date{Accepted XXX. Received YYY; in original form ZZZ}

\pubyear{2019}

\begin{document}
\label{firstpage}
\pagerange{\pageref{firstpage}--\pageref{lastpage}}
\maketitle

\begin{abstract}
The observable properties of galaxies depend on both internal processes and the external environment.  In terms of the environmental role, we still do not have a clear picture of the processes driving the transformation of galaxies.  The use of proxies for environment (e.g.,  host halo mass, distance to the N$^{\rm th}$ nearest neighbour, etc.), as opposed to the real physical conditions (e.g., hot gas density) may bear some responsibility for this.  Here we propose a new method that directly links galaxies to their local environments, by using spatial cross-correlations of galaxy catalogues with maps from large-scale structure surveys (e.g., thermal Sunyaev-Zel'dovich [tSZ] effect, diffuse X-ray emission, weak lensing of galaxies or the CMB).  We focus here on the quenching of galaxies and its link to local hot gas properties.  Maps of galaxy overdensity and quenched fraction excess are constructed from volume-limited SDSS catalogs, which are cross-correlated with tSZ effect and X-ray maps from Planck and ROSAT, respectively.  Strong signals out to Mpc scales are detected for most cross-correlations and are compared to predictions from the EAGLE and BAHAMAS cosmological hydrodynamical simulations.  The simulations successfully reproduce many, but not all, of the observed power spectra, with an indication that environmental quenching may be too efficient in the simulations.   We demonstrate that the cross-correlations are sensitive to both the internal (e.g., AGN and stellar feedback)  and external processes (e.g., ram pressure stripping, harassment, strangulation, etc.) responsible for quenching.  The methods outlined in this paper can be adapted to other observables and, with upcoming surveys, will provide a stringent test of physical models for environmental transformation.
\end{abstract}




\section{Introduction}

Surveys such as the Sloan Digital Sky Survey (SDSS) and Galaxy and Mass Assembly (GAMA) (among others) have shown that in the local Universe, galaxies can be classified into two broad populations: star-forming and passive. Star-forming systems typically have blue colours \citep{Strateva_etal_2001, Blanton_etal_2003, Baldry_etal_2004}, late-type morphologies \citep{Wuyts_etal_2011, van_der_Wel_etal_2014}, young stellar populations \citep{Kauffmann_etal_2003, Gallazzi_etal_2008}, and high star formation rates (SFR) \citep{Noeske_etal_2007, McGee_etal_2011, Wetzel_2012}. On the other hand, passive galaxies exhibit red colours, early-type morphologies, old stellar ages $-$ all of which can be connected to their low star formation rates.

These effects manifest themselves in the form of a well-known galaxy bimodality when plotted as a function of stellar mass, persisting out to redshifts as high as $\mathrm{z \sim 4}$ \citep{Baldry_etal_2006, Brammer_etal_2009, Muzzin_etal_2013}. There is a tight, positive correlation for blue, star-forming galaxies known as the `main sequence' (MS), which is in stark contrast to the passive population dominating the high-stellar mass end. The presence of a bimodality suggests that a relatively rapid transition occurs in the evolutionary sequence of a galaxy where it ceases its star formation, i.e. quenching. 

It is yet undetermined which exact physical processes are responsible for this evolution.  It is possible, however, to separate them out into two categories: internal and external processes \citep{Peng_etal_2010}. The former, secular mode, which can occur in all galaxies irrespective of external factors, is strongly correlated with stellar mass \citep{Driver_etal_2006, Barro_etal_2017}. Indeed, this mode is more pronounced at $\mathrm{\log_{10}(M_*/M_{\odot})} \gtrsim 10$, where Active Galactic Nucleus (AGN)-driven outflows \citep{Nandra_etal_2007} and heating suppress star formation processes.  At the low-mass end, stellar feedback is thought to regulate star formation activity. In galaxies with $\mathrm{\log_{10}(M_*/M_{\odot})} \lesssim 9$, which do not host a strong AGN, stellar-driven outflows eject the more loosely-bound gas \citep{DellaVechia_etal_2008} but may not remove it completely, allowing it to fall back onto the galaxy where it can become available for further star formation.

A galaxy's local environment can lead to additional quenching by inhibiting the supply or outright removing gas required to fuel star formation.  It is now well established that over-dense environments host galaxies with suppressed star formation rates and the observational features outlined above \citep{Oemler_1974, Postman_Geller_1984, Dressler_etal_1999, Blanton_Moustakas_2009, Kimm_etal_2009, Peng_etal_2010, Wetzel_2012}.  Even on an individual cluster scale, gradients in quenched fraction ($\mathrm{f_q}$) have been observed to correlate with cluster-centric distance \citep{Rasmussen_etal_2012, Wetzel_2012, Haines_etal_2015, Barsanti_etal_2018}. In addition, satellite galaxies are observed to be more metal-rich in their ionised gas and stars \citep{Pasquali_etal_2012, Bahe_etal_2017, Maier_etal_2019}. Relative to the field, enhanced quenched fractions extending beyond the virial radii of clusters have also been observed and attributed to `pre-processing', whereby galaxies begin quenching as part of smaller groups prior to in-falling onto a cluster \citep{Fujita_2004, Lu_etal_2012, Wetzel_2012, Bahe_etal_2013, Roberts_etal_2017}.

There are many possible mechanisms for quenching satellite galaxies as they spiral in towards massive groups and clusters, either individually or as part of a smaller group.  For example, the cold, star-forming gas of infalling galaxies can be directly ram pressure stripped (RPS) due to the relative motion of the galaxy with respect to diffuse medium of the host system \citep{Gunn_Gott_1972, Abadi_etal_1999, Quilis_etal_2000, Poggianti_etal_2017, Brown_etal_2017, Barsanti_etal_2018}.  Material being circulated in feedback-driven galactic fountains may be even more susceptible to RPS \citep{Bahe_etal_2015}.  Turbulent viscous stripping \citep{Nulsen_1982, Kraft_etal_2017} acts as an additional form of stripping alongside RPS.  `Strangulation' (or `starvation') refers to the process through which the hot gas reservoir of an infalling galaxy, which supplies the fuel for ongoing star formation, is removed \citep{Larson_etal_1980, Moore_etal_1999, Balogh_etal_2000, Kawata_Mulchaey_2008, McCarthy_etal_2008, Peng_etal_2015}.  In this case, the infalling galaxy is quenched more slowly, on an ISM gas consumption time scale.  

RPS, turbulent viscous stripping, and strangulation represent hydrodynamical processes.  However, gravitational interactions can also result in the quenching of galaxies.  For example, galaxy-galaxy mergers can cause an initial burst of star formation but leave the galaxy quenched in the end \citep{Mihos_Hernquist_a_1994, Mihos_Hernquist_b_1994, Schawinski_etal_2014}.  `Harassment'  is a less dramatic process of inducing starbursts through repeated dynamical interactions \citep{Farouki_Shapiro_1981, Moore_etal_1996, Hirschmann_etal_2014}, thus exhausting the gas reservoir.  In addition, tidal interactions with the overall group/cluster potential well \citep{Mayer_etal_2006, Chung_etal_2007} can disturb the cooling and accretion of gas making it less bound and, therefore, easier to be stripped by other quenching modes.

In addition to the above hydrodynamical and gravitational processes, it is possible that the local radiation (e.g., \citealt{Kannan_etal_2016}) and magnetic fields (e.g., \citealt{Tonnesen_etal_2014}) may also play a role in the quenching of satellites.  However, at present these possibilities are ill-constrained by observations.

The main limitation in our inability to robustly identify the main mechanism(s) behind environmental quenching (and its possible dependence on, e.g., time or galaxy mass) likely stem from our inability to observe each process independently.  Instead, the focus has shifted to characterising when/where (rather than how/why) galaxies become quenched and then trying to use this information to test different physical models.  For example, the currently favoured scenario is the `delayed-then-rapid' scenario proposed by \citet{Wetzel_etal_2013}. This is a two-stage process, whereby a satellite galaxy experiences a slow form of quenching as it initially falls into a cluster (possibly through starvation). As it reaches the dense, inner part of the ICM, a short-timescale process (such as RPS) becomes much more efficient and begins to dominate - inducing a rapid form of quenching. Evidence for this two-stage scenario comes from studying quenching timescales \citep{Muzzin_etal_2014, Peng_etal_2015, Fillingham_etal_2015, Foltz_etal_2018} or observing RPS in action through extended HI distributions around satellite galaxies in dense environments \citep{Kenney_etal_2015, Poggianti_etal_2016, Jaffe_etal_2016}.

The above-mentioned studies focussed on observing galaxy properties as a function of environment, i.e. an over-density probed via the same galaxies. There are a number of approaches used in estimating the local density; for example, group membership counts from a catalogue such as \citet{Yang_etal_2007}, nearest neighbour distance \citep{Park_etal_2007}, number counts of neighbouring galaxies in a defined volume (usually cylindrical) \citep{Kauffmann_etal_2004, Blanton_Moustakas_2009}, or taking their velocity into account as well through phase space diagrams \citep{Pimbblet_etal_2006, Oman_Hudson_2016}. Some studies have investigated environmental effects via angular cross-correlations, asking the question of whether galaxies are clustered differently when separated by their star formation rate \citep{Hatfield_Jarvis_2017} or morphology \citep{Cervantes_etal_2016}. However, while it is true that galaxy overdensity is a tracer of the underlying overdensity and environmental quenching correlates with it, probing the overdensity stands little chance of determining \emph{which} of the processes are more dominant. This is because the majority of the different cluster mechanisms/components responsible for quenching correlate with the underlying matter density distribution. It is, therefore, necessary to observe multiple components at the same time and measure the correlation with galaxy properties in order to break the degeneracies present.   This is easier said than done, however, as the diffuse ICM is extremely faint in X-ray emission and unbiased estimates of total overdensity from weak lensing are noisy as well as low in number, particularly on the scale of groups.

One recent attempt has been made using deep \emph{Chandra} X-ray observations of low-redshift clusters in SDSS by \citet{Roberts_etal_2019}.  The authors find evidence for a threshold in ICM density which separates regions of gradually-increasing quenched fraction and sudden steepening of the trend closer to the cluster centre.  This is interpreted as the dominant quenching mechanism transitioning from a steady gas depletion (e.g. starvation) to a more abrupt gas removal process like direct ram pressure stripping of the ISM.

In this study, we introduce a new test of environment based around spatial cross-correlations between observables of the ICM/gravitational potential and large-area galaxy catalogues. This map-based approach offers the advantages of extracting a signal from otherwise low signal-to-noise observations by measuring over a large area of the sky, avoiding the complex process of finding groups and clusters, and potentially breaking the degeneracies between different cluster components.  Such methods have already been employed in other areas of astrophysics, such as galaxy$-$CMB lensing to probe cosmology via the growth of structure \citep{Giannantonio_etal_2016}, thermal Sunyaev-Zel'dovich (tSZ) effect $-$ X-ray emission $-$ CMB weak lensing in order to measure halo bias in the clustering of dark matter \citep{Hurier_etal_2019}, and the unresolved $\mathrm{\gamma}$-ray background $-$ galaxy cluster cross-correlation to study the nature of this $\mathrm{\gamma}$-ray emission \citep{Hashimoto_etal_2019}. As a first application of the method to environmental quenching, we focus on the hot gas$-$quenched fraction signal.  We construct large-area maps of galaxy overdensity and quenched fraction from the SDSS spectroscopic and photometric surveys, compute auto- and cross-power spectra between galaxy survey quantities and two measures of ICM.  To characterise the state of the hot gas, we make use of maps of the thermal Sunyaev-Zel'dovich (tSZ) effect as measured by the $Planck$ satellite and X-ray emission from the ROSAT All-Sky Survey (RASS), respectively.  We perform equivalent measurements on synthetic maps produced from state-of-the-art hydrodynamical simulations: EAGLE \citep{Schaye_etal_2015,Crain_etal_2015, McAlpine_etal_2016} and BAHAMAS \citep{McCarthy_etal_2017,McCarthy_etal_2018}. 

This paper is organised in the following sections: we describe the galaxy catalogues, X-ray, and tSZ effect data and map-making procedures, as well as simulations in this study, in Section~\ref{section:data}.   The formalism of cross-correlating two discretised maps is outlined in Section~\ref{section:cross-correlation}.  Our main results are presented in Section~\ref{section:results} and discussed further in Section~\ref{section:discuss}.  Finally, we summarise our findings in Section~\ref{section:conclusions}.

Throughout, we adopt a flat $\Lambda$CDM concordance cosmology with $\mathrm{\Omega_{m}} = 0.274$ and $H_0 = 70.5~\mathrm{km/s/Mpc}$ \citep{Hinshaw_etal_2009}.

\section{Data and Map-making}
\label{section:data}
\subsection{Galaxy catalogues}
For our galaxy samples we use the Sloan Digital Sky Survey (SDSS) DR7 \citep{Abazajian_etal_2009} and DR12 \citep{Alam_etal_2017} to construct two volume-limited samples.  As described below, the DR7 is used to construct a shallower spectroscopic sample ($z < 0.06$), while the DR12 is used to construct a deeper ($z < 0.15$) photometric sample.

\subsubsection{SDSS DR7 spectroscopic sample}
The MPA-JHU\footnote{https://wwwmpa.mpa-garching.mpg.de/SDSS/DR7} value-added galaxy catalogue \citep{Brinchmann_etal_2004} provides derived galaxy properties from emission line analysis of SDSS DR7. We use their stellar masses, specific star formation rates ($\mathrm{sSFR~=~SFR/M_*}$), observed \texttt{cModel} magnitudes, redshifts, and positions, and call this the `spectroscopic' sample. 
To form our sample, we first select all objects identified as galaxies with the \verb|TARGETTYPE = GALAXY| parameter in the \verb|gal_info_dr7_v5_2.fits| file. Next, we select all galaxies with reliable (specific) star formation rates, i.e. \verb|FLAG = 0| in the \verb|gal_totspecsfr_dr7_v5_2.fits|. The catalogue is complete to Petrosian r-band magnitude $r \leq 17.77$ over most of the SDSS footprint, but to ensure a consistent sky coverage we use a more conservative limit of $r \leq 17.5$.  Note that \texttt{cModel} and Petrosian magnitudes are sufficiently similar so that we can ignore the differences between them in terms of selection.  Post selection we use the former as it provides a more reliable estimate of a galaxy's total flux and has close to optimal noise properties \citep{Stoughton_etal_2002}.

In order to aid the interpretation of the observations and to make a straightforward and consistent comparison to simulations (described in Section 2.6), we opt for a volume- and stellar mass-complete sample.  Ideally, we would like to probe the low-mass end of the galaxy population as these galaxies are more likely to be quenched due to environmental effects.  With a flux-limited survey, however, a balance must be struck between the redshift limit and the lower mass limit if a volume-limited sample is desired.  A lower stellar-mass limit is also introduced by simulations, as the resolution is finite and the low-mass galaxy properties become unreliable. BAHAMAS is the lower resolution simulation of the two used in this study and has been shown to reproduce galaxy properties down to a stellar mass of $\mathrm{log}_{10}[M_*/\mathrm{M_{\odot}}]=10$.   We therefore adopt this as our lower-limit in galaxy stellar mass.

To determine the limiting redshift for a given stellar mass cut in a volume-limited sample, we adopt the method described in \citet{Baldry_etal_2018}.  Specifically, we examine the mass-to-light ratio in the i-band ($M_*/L_i$) against stellar mass ($M_*$) in a redshift slice; a clear drop-off in $M_*/L_i$ can be seen for stellar masses which are no longer completely sampled given the r-band limit of the survey. We adjust our redshift upper limit such that the drop-off occurs at $\mathrm{log}_{10}[M_*/\mathrm{M_{\odot}}]$ just below 10. For a volume-limited sub-sample (selected from the main SDSS sample) of galaxies with $\mathrm{log}_{10}[M_*/\mathrm{M_{\odot}}]>10$, the upper limit in redshift is $z=0.06$.  If a higher-redshift/larger volume is desired, the lower limit on stellar mass must be raised for the sample to remain volume-complete.  But since the aim of this study is to characterise environment, it is necessary to cover the range in $M_*$ over which galaxies transition from blue/star-forming to red/quenched, so a lower stellar mass limit is preferred to one that probes larger volumes but with higher-mass systems.

We further split our sample into star-forming and quenched sub-samples by introducing a simple division in sSFR following \citet{Wetzel_2012} at $\mathrm{log_{10}[sSFR (\mathrm{yr^{-1}})]} = -11$. This division is then used to compute quenched fraction in Section \ref{section:map-making}.

\subsubsection{SDSS DR12 photometric sample}
In order to obtain a larger and deeper sample (in terms of limiting redshift) of galaxies, we also use a sample not restricted by the spectroscopic completeness of SDSS. For this purpose we use SDSS DR12 data, with photometric redshifts and parametrically-estimated stellar masses and refer to it as the `photometric' sample.

We use photometric redshift estimates of \citet{Beck_etal_2016} obtained using a hybrid method of machine learning and template-fitting techniques. They achieve a normalised mean redshift estimation error of $\mathrm{\overline{\Delta z_{norm}}} = 5.84\times 10^{-5}$, where 
$\mathrm{\Delta z_{norm} = (z_{phot} - z_{spec}) / (1 + z_{spec})}$, a standard deviation $\mathrm{\sigma(\Delta z_{norm})} = 0.0205$, and an outlier rate of $\mathrm{P_o} = 4.11\%$.
Since spectroscopic redshifts are available for all galaxies with $\mathrm{r} \lesssim 17.77$ in the main SDSS footprint (we use these in the spectroscopic sample), and photo$z$ errors have the largest impact at low redshift, we use MPA-JHU redshift estimates where available. Therefore, although the sample is called `photometric', the redshifts used are a combination of spec$z$ and photo$z$.   The stellar masses are estimated using the same method (described below) for all galaxies, independent of whether their redshifts are spectroscopic or photometric, in order to stay consistent throughout the sample.

Empirical stellar mass estimates are computed following \citet{Sedgwick_etal_2019}, which itself is based on the method outlined in \citet{Taylor_etal_2011} and \citet{Bryant_etal_2015}, and calibrated using SED-fitting data from the GAMA \citep{Baldry_etal_2018} survey. The estimation relies on the correlation between mass-to-light ratio and colour.  One can write down an equation for stellar mass that depends only on distances, redshifts, and observed magnitudes and which folds in the k-correction:
\begin{equation}
\mathrm{log}_{10}(M_*/\mathrm{M_{\odot}}) = -0.4i + 0.4D +f(z) + g(z)(g-i)_{obs},
	\label{eq:stellar_mass}
\end{equation}
\noindent where $i$ is the i-band observed \texttt{cModel} apparent magnitude, $D$ is the distance modulus to the galaxy, $z$ is the redshift estimate, $(g-i)_{obs}$ is the observed $g-i$ colour\footnote{The colour is computed using \texttt{model} magnitudes in order to stay consistent between the two bands, as the aperture parameters are determined in the r-band and applied to all other bands.}, and $f(z)$ and $g(z)$ are fitted polynomial functions of redshift:
\begin{equation}
\begin{split}
&f(z) = -15.15z^3 + 9.193z^2 - 1.687z + 1.104,\\
&g(z) = 26.40z^3 - 12.84z^2 + 0.5908z + 0.8237.
\end{split}
	\label{eq:f_and_g}
\end{equation}

\begin{figure}
	\includegraphics[width=\columnwidth]{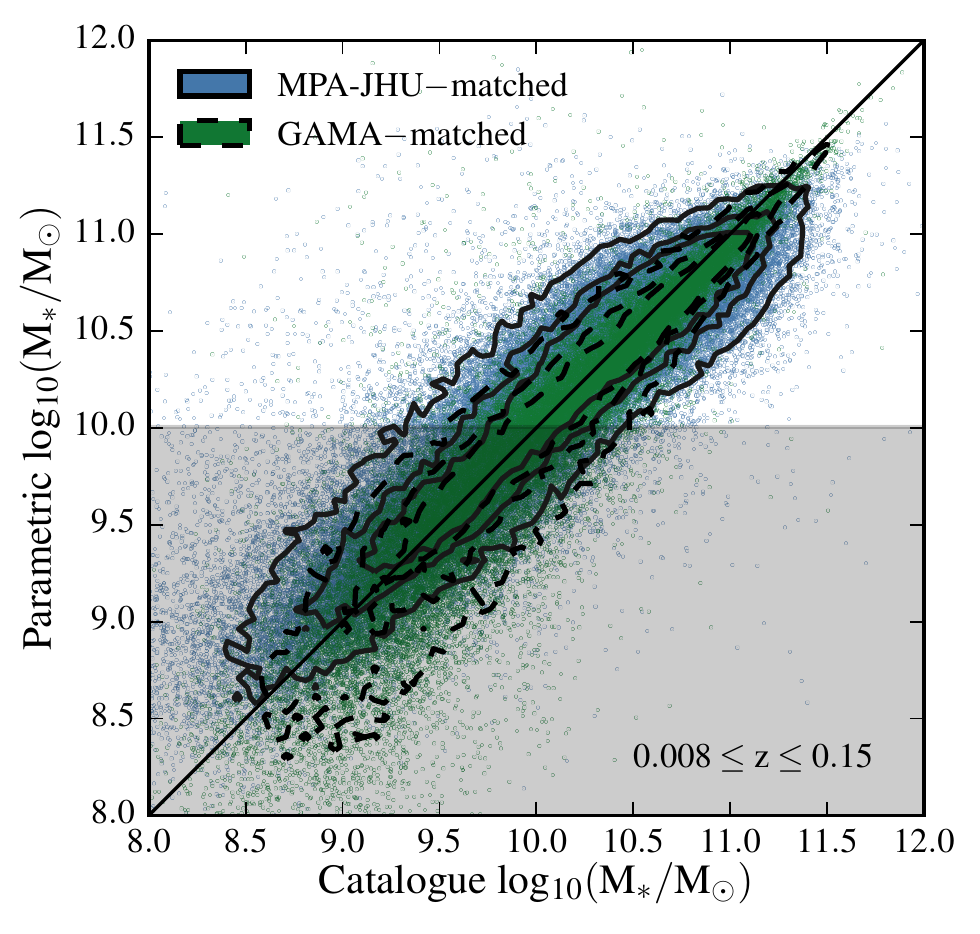}
    \caption{A comparison of parametric stellar mass estimates using equation~\ref{eq:stellar_mass} to MPA-JHU estimates using line ratios, and GAMA SED-derived estimates. Blue with solid, black contours shows MPA-JHU$-$matched galaxies, whereas green with dashed contours corresponds to GAMA$-$matched sample; they represent 90th, 70th, and 50th percentiles. Galaxies in the photometric sample are matched to both catalogues by their SDSS object ID.  Scatter in the stellar mass estimates is $\sim0.3$ dex but there is no significant bias in the stellar mass estimates of the photometric sample.}
    \label{fig:stellar_mass_check}
\end{figure}

To test the derived stellar masses, we match our photometric sample galaxies to MPA-JHU and GAMA\footnote{The GAMA DR3 footprint is, unfortunately, too small to achieve significant detections of the cross-correlations with the tSZ effect and X-ray data used in this study.  Future high sensitivity X-ray (e.g., with eROSITA) and tSZ effect data (e.g., Simons Observatory, CMB-S4) will allow this issue to be overcome.  Larger deep spectroscopic surveys (e.g., WAVES, DESI) are also expected to improve these detections dramatically.  In the present study, we therefore use GAMA solely for the purpose of calibrating stellar mass estimates of the SDSS photometric sample.} catalogues by \texttt{ObjID} and compare catalogue (i.e., spectra-based) stellar masses to those resulting from eqn.~\ref{eq:stellar_mass}. This can be seen in Figure~\ref{fig:stellar_mass_check}. A comparison to MPA-JHU tests the parametric fit without photo$z$ errors (spectroscopic redshifts were used for everything with $\mathrm{r} \leq 17.5$) and matching to GAMA galaxies tests the validity of the estimate overall. From Fig.~\ref{fig:stellar_mass_check} it is apparent that while there is scatter of $\sim 0.3~\mathrm{dex}$ in the empirically-derived stellar masses, there does not appear to be any systematic bias. Furthermore, the visibly larger scatter at $\mathrm{log_{10}}[M_*/\mathrm{M_\odot}] < 10$ has no consequence for our sample as we exclude these (grey shaded region) galaxies. Scatter across the boundary, i.e. contamination, is at the $\sim 5 \%$ level and is approximately equal in both directions. Therefore, we conclude that photometric redshifts combined with empirical stellar mass estimates do not significantly bias our galaxy samples.
When selecting our photometric sample, we adopt the r-band Petrosian magnitude limit of $r < 19.8$ from the GAMA DR3 survey, making the assumption that the rest of SDSS field is complete at least down to this magnitude, and use the same redshift limit of $z \leq 0.15$ for a volume-complete sample of galaxies with $\mathrm{log_{10}}[M_*/\mathrm{M_{\odot}}] \geq 10$ from \citet{Baldry_etal_2018}.

\begin{figure}
	\includegraphics[width=\columnwidth]{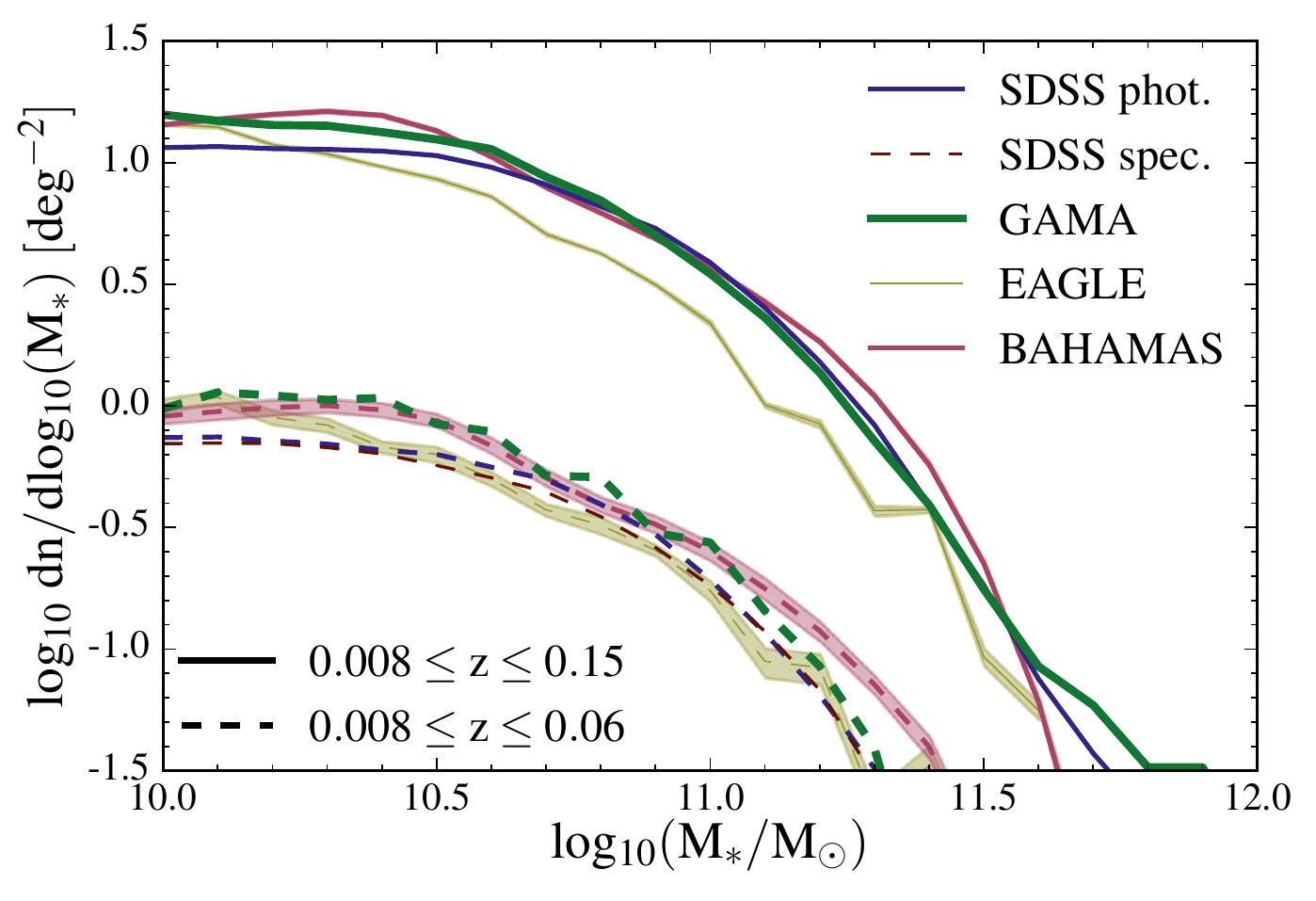}
    \caption{Galaxy stellar mass functions for both samples, normalised by their respective footprint area. Solid and dashed curves correspond to the spectroscopic and photometric samples, respectively.  The SDSS photometric sample slightly underestimates the GSMF at masses below $\sim 10^{10.7} \mathrm{M_\odot}$ with respect to GAMA, achieving a good match otherwise.  For completeness, a low-redshift version of photometric (dashed, blue curve) is shown, which is nearly identical to SDSS spectroscopic sample.  Also shown are the GSMFs from EAGLE and BAHAMAS, which were calibrated on previous estimates of the GSMF.  Shaded regions represent the $1\sigma$ confidence interval on the mean value of 10 simulated light cones.  Overall the simulations yield similar GSMFs to those derived from our observational samples, although EAGLE tends to fall somewhat below the observations at the knee of the mass function (as also found previously). }
    \label{fig:stellar_mass_fn}
\end{figure}

As an additional test of our samples, we plot in Fig.~\ref{fig:stellar_mass_fn} the galaxy stellar mass functions (GSMF) for the two SDSS samples and compare them to that derived from GAMA and from the simulations used in this study.  Here we define the galaxy stellar mass function as the number of galaxies per decade in stellar mass per unit angular area on the sky, within the two redshift limits mentioned above.  (The survey angular area is obtained from the \texttt{HEALPix}\footnote{http://healpix.sourceforge.net} maps described below.  For reference, the survey areas are: $\mathrm{SDSS} = 7849$, $\mathrm{GAMA} = 153$, $\mathrm{EAGLE} = 100$, $\mathrm{BAHAMAS} = 625$ square degrees.) Dashed curves are for the spectroscopic selection ($z<0.06$) and solid curves for photometric selection ($z<0.15$). 

Good consistency is obtained for both the SDSS photometric and spectroscopic GSMFs with those derived from GAMA.  For completeness, although this sample is not used in any further analysis, we also show a low-redshift version of the photometric sample ($z<0.06$), which yields a near identical GSMF to that derived from the spectroscopic sample.  This suggests that scatter seen in Fig.~\ref{fig:stellar_mass_check} does not affect the stellar mass distribution statistics in a significant way.  Close agreement is also achieved with the simulations.  This is not particularly surprising, as the feedback prescriptions in both EAGLE and BAHAMAS were tuned to reproduce estimates of the local GSMF from previous studies (see \citealt{Schaye_etal_2015,McCarthy_etal_2017}).  However, it is reassuring that neither our sample selection and stellar mass estimation on the observation side, nor our light cone making methods on the simulation side, introduce systematic biases.  In the case of EAGLE, the simulations fall somewhat below the observations near the knee of the mass function for the deeper photometric selection.  This is consistent with what was found previously in \citet{Schaye_etal_2015} (see their figure~4).  

\begin{figure}
	\includegraphics[width=\columnwidth]{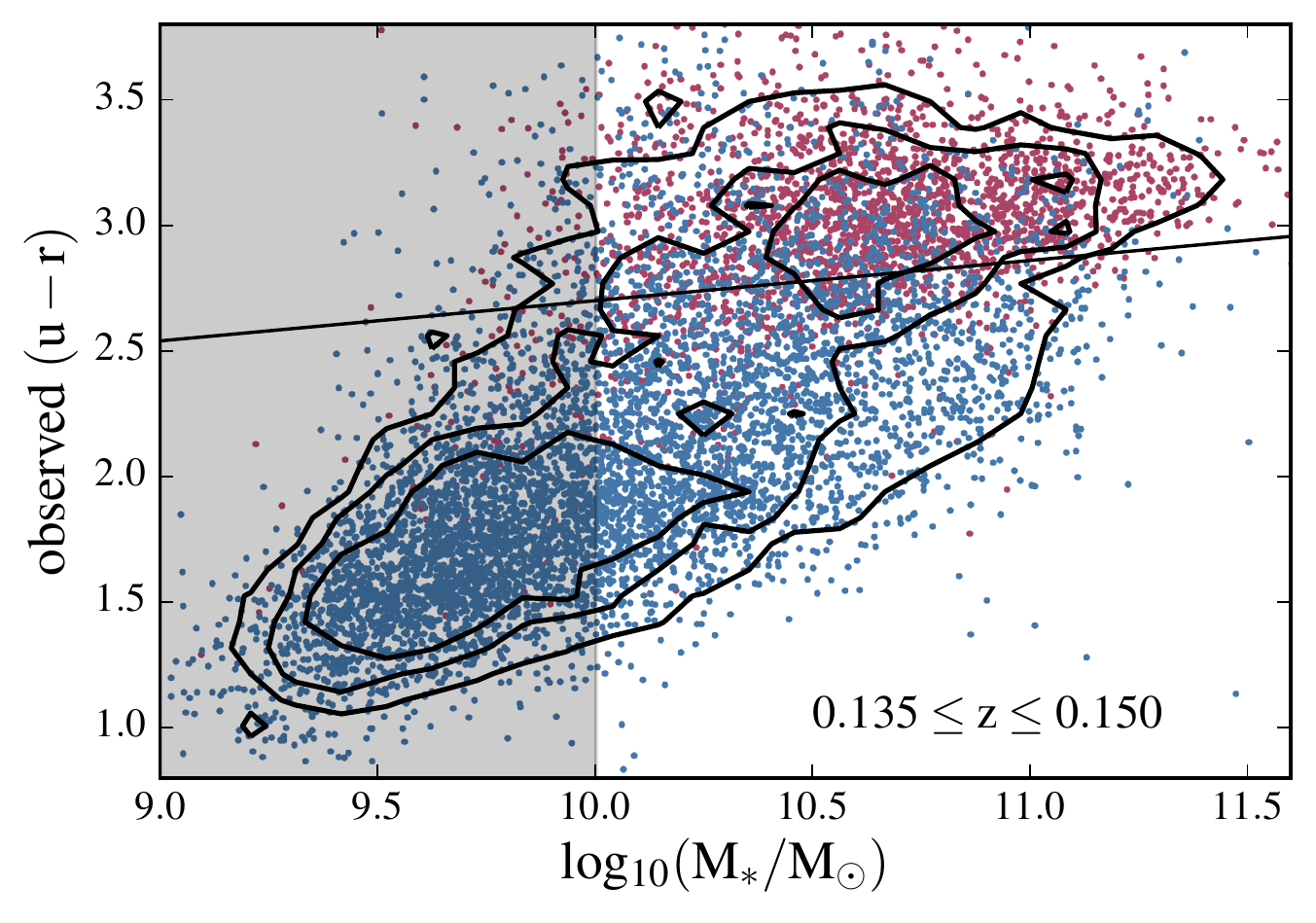}
    \caption{An example of the colour$-$stellar mass relations used to calibrate the quenched$-$star-forming division for the SDSS photometric sample. SDSS galaxies are matched to GAMA and assigned a binary (red/blue) sSFR flag.   Virtually all quenched galaxies (on the basis of their GAMA sSFR) lie on the `red sequence'.  Some star-forming galaxies also lie on the red sequence (presumably as a result of strong dust reddening, see text).  The black solid line shows the colour division which achieves the same mean quenched fraction and $\mathrm{f_q-M_*}$ relation as for GAMA (see Fig.~\ref{fig:fq_logMstar}). The shaded region marks stellar masses which fall below our limit for volume-completeness.}
    \label{fig:colour_ssfr_correlation}
\end{figure}

\begin{figure}
	\includegraphics[width=\columnwidth]{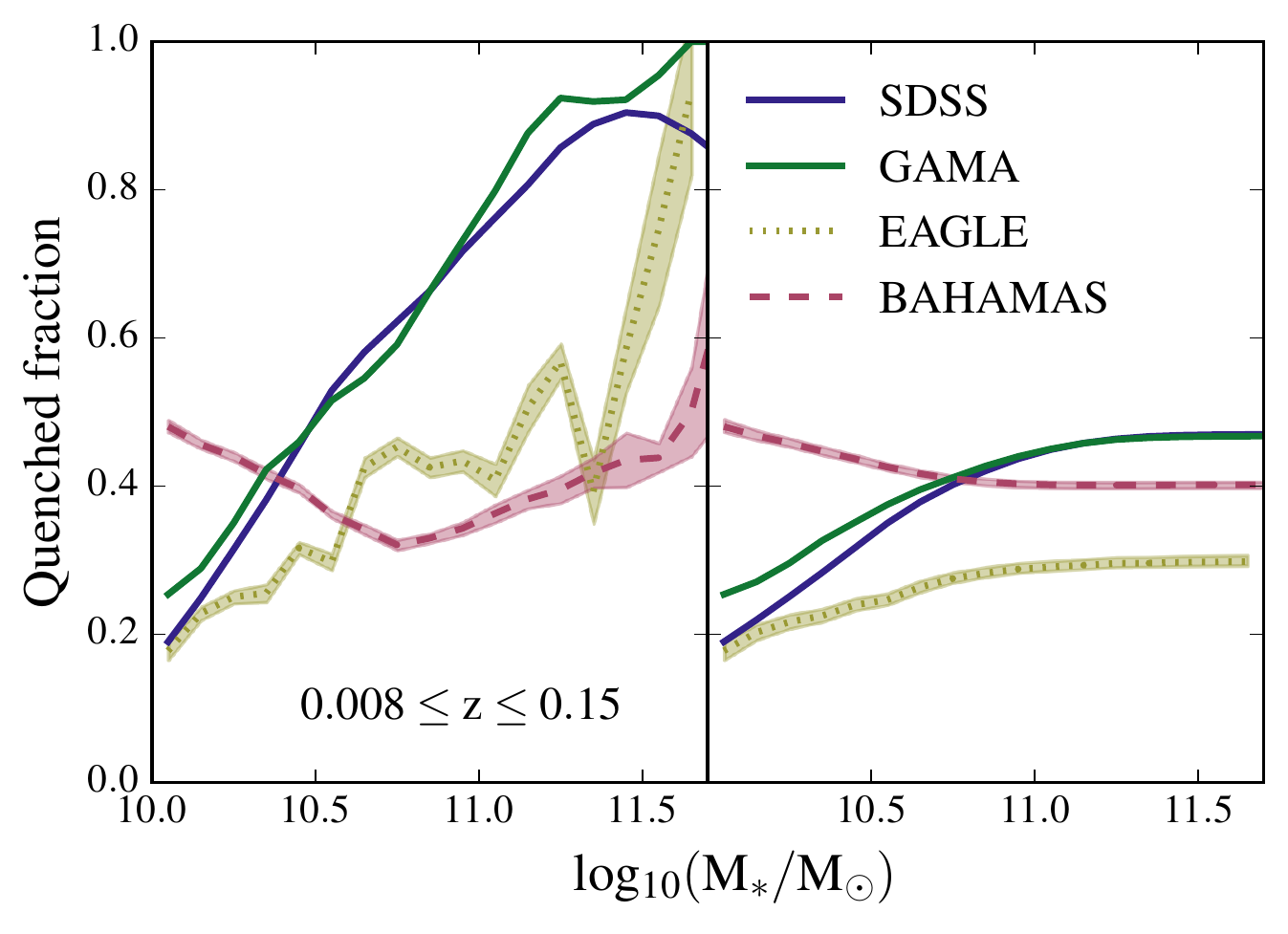}
    \caption{\emph{Left}: Quenched fraction ($\mathrm{f_q}$) as a function of stellar mass for the $0.008 \leq z \leq 0.15$ sample. $\mathrm{f_q}$ is computed in each $\mathrm{log_{10}(M_*)}$ bin using sSFR for GAMA, EAGLE, and BAHAMAS, and $\mathrm{(u-r)_{obs}}$ colour for SDSS.  The solid and dashed curves represent the observations and simulations, respectively, with shaded regions showing the $1\sigma$ scatter regions around the median of 10 light cones for the simulations. \emph{Right}: Cumulative quenched fraction, $f_q(<\mathrm{M_*})$, as a function of stellar mass.  The SDSS photometrically-derived trend matches the spectroscopically-inferred relation from the smaller GAMA calibration set.  The simulations predict quenched fractions that fall significantly below what is observed for stellar masses of $\mathrm{log_{10}[M_*/M_\odot]} > 10.5$.  However, galaxies near the lower limit $\mathrm{log_{10}[M_*/M_\odot]} = 10.0$ dominate the sample (see text). }
    \label{fig:fq_logMstar}
\end{figure}

With no estimates of the SFR for galaxies without spectra, we cannot use sSFR as a means to separate star-forming and quenched galaxies for the photometric sample.  Instead, we use the observed $\mathrm{(u-r)}$ colour--$\mathrm{M_*}$ relation to divide the sample. In Fig.~\ref{fig:colour_ssfr_correlation} we show this relation for GAMA-matched galaxies in the photometric sample, where galaxies are coloured according to quenched status (red=quenched, blue=star-forming) based on their sSFR.  It is clear that most of the galaxies belonging to the `red sequence' are quenched, as determined from their sSFRs.  However, there is also a population of star-forming galaxies which inhabit the red sequence, presumably as a result of strong reddenning by dust \citep{Evans_etal_2018}.  We demonstrate later that this small contamination by star-forming galaxies is negligible for our cross-correlations.

Given that quenched galaxies lie almost exclusively on the red sequence, we can use a galaxy's colour to assign a quenched flag. We do this by first sub-dividing the sample into three redshift bins with approximately equal number of galaxies in each. This is done in order to reduce the effect of k-corrections and account for any possible evolution of colour with redshift in the chosen range (although $0.008 \leq z \leq 0.15$ is small enough so that evolutionary effects are not significant).  Next, a line is fitted to the red sequence in the GAMA-matched sub-sample to obtain a slope and intercept.  We apply this relation to the full SDSS photometric sample, maintaining the slope of the relation but adjusting the intercept until the mean quenched fraction matches that of the GAMA-matched sample (computed using sSFR) in each bin. The process is repeated for all three redshift bins, visually inspecting the $\mathrm{(u-r)}$ colour--$\mathrm{M_*}$ relation, and assigning a quenched flag to each galaxy.

The resulting distribution is verified in the left-hand panel of Figure~\ref{fig:fq_logMstar}, which shows the relation between quenched fraction and stellar mass.  To within $50\%$ the colour-determined $\mathrm{f_q}$--$\mathrm{M_*}$ relation derived from the SDSS photometric sample matches the sSFR-determined $\mathrm{f_q}$--$\mathrm{M_*}$ relation derived from the smaller GAMA spectroscopic calibration sample.

Computing $\mathrm{f_q}$ for the simulations (using SFR and $\mathrm{M_*}$ computed within a $30~\mathrm{kpc}$ aperture) yields relations which fall significantly below what is observed for stellar masses of $\mathrm{log_{10}[M_*/M_\odot]} > 10.5$.  (The result for EAGLE is consistent with that shown previously by \citealt{Schaye_etal_2015} and \citealt{Furlong_etal_2015}.)  Here we note that the different feedback schemes employed in EAGLE and BAHAMAS were not calibrated on this metric and were therefore not guaranteed to reproduce these observations.  Nevertheless, this comparison illustrates that there are still some deficiencies in the feedback prescriptions of these simulations.  Na\"ively, one might expect these deficiencies to compromise comparisons of cross-correlations involving quenched fraction.  However, it is important to note that for the selection employed in this study, the vast majority of the signal is dominated by galaxies near the lower stellar mass limit, where the simulations have reasonable quenched fractions.  This is just by virtue of the fact that the lower mass galaxies dominate the sample by abundance.  To illustrate this, in the right hand panel of Figure~\ref{fig:fq_logMstar} we show the cumulative quenched fraction as a function of stellar mass.  This relation reveals that galaxies above $\mathrm{log_{10}[M_*/M_\odot] \sim 10.7}$ contribute very little to the total number of galaxies in the sample.  We have also checked that the cross-correlations we present later (in Section \ref{section:results}) do not qualitatively change when we exclude high-mass galaxies with $\mathrm{log_{10}[M_*/M_\odot] > 10.5}$ from our samples\footnote{Given the differences at the high-mass end present in the left-hand panel of Figure~\ref{fig:fq_logMstar}, we could expect significant differences between the simulations and observations for cross-correlations involving exclusively high-mass systems.  However, the relatively low abundance of high-mass systems results in noisy estimates of these cross-correlations at present.  Deeper observations (e.g., with DES, LSST, Euclid) will resolve this issue in the near future.}.  We also point out that the auto- and cross-power spectra that we examine involve overdensities with respect to the mean quenched fraction, rather than the mean quenched fraction itself.

Several basic sample properties are summarised in Table~\ref{table:samples}.

\begin{table}
\begin{centering}
\begin{tabular}{c|c|c|}
\cline{1-3}
\textbf{}               & \textbf{Spectroscopic} & \textbf{Photometric} \\ \cline{1-3} 
r-band limit            & 17.5                   & 19.8                 \\
$\mathrm{M_{*,min}}$                & $10^{10} {\rm M_\odot}$  & $10^{10} {\rm M_\odot}$ \\  
$\mathrm{z_{min}}$                  & 0.008                  & 0.008                \\
$\mathrm{z_{max}}$                  & 0.06                   & 0.15                 \\
$\mathrm{z_{med}}$                  & 0.047                  & 0.118                 \\
$\mathrm{N_{gal}}$                  & 44799                  & 953980    \\   
$\mathrm{\overline{f_q}}$\tablefootnote{These values are final mean quenched fractions of the maps after all masks have been applied.}           & 0.418                  & 0.494        \\ \cline{1-3} 
\end{tabular}
\caption{A summary of galaxy sample properties for the spectroscopic and photometric SDSS volume-limited samples.}
\label{table:samples}
\end{centering}
\end{table}

\subsection{\texttt{HEALPix} map-making using galaxy catalogues}
\label{section:map-making}

In order to perform the cross-correlations between galaxy and hot gas properties we need to construct equivalent maps for the properties we are interested in.  In terms of galaxy properties, in this paper we focus on two quantities: total galaxy overdensity ($\mathrm{\breve{N}_{tot}}$), and quenched fraction ($\mathrm{\Breve{f}_{q}}$).  We adopt the `breve' (` $\breve{}$ ') notation to denote excess-relative-to-the-mean quantities, such as overdensity: $\mathrm{\breve{N}_{tot}}~=~(\mathrm{N_{tot}}~-~\mathrm{\bar{N}_{tot}})/\mathrm{\bar{N}_{tot}}$,
where $\mathrm{N_{tot}}$ is the total projected galaxy surface density and $\mathrm{\bar{N}_{tot}}$ is its mean value. This ensures that maps for both galaxy measures are in the same $[-1, \infty)$ range, and have a mean value of zero.

For the galaxy-based maps, we adopt the same \texttt{HEALPix} \citep{Gorski_Hivon_2011} format and resolution ($\sim1.716$ arcmin, $\mathrm{N_{SIDE}=2048}$) as used for the tSZ effect and X-ray maps, which are described below.  To compute the cross-correlations (described in Section \ref{section:cross-correlation}), we use tools (e.g., \verb'NaMASTER') originally designed for analysis of contiguous fields in the \texttt{HEALPix} format, such as those regularly produced using cosmic microwave background data.  The \verb'NaMASTER' algorithm has the capability to deal with non-contiguous/incomplete fields to an extent, but we have found that even for our larger photometric sample, the galaxies are too sparsely distributed for the algorithm to give reliable results if we simply mask empty pixels.  For galaxy density, this can be overcome by filling the empty pixels (within SDSS footprint) with a zero value, while masking everything outside the main footprint.  However, this solution will not work for the quenched fraction, as zero-valued pixels would represent fully star-forming regions, and masking is not an option for already mentioned reasons. We therefore employ adaptive smoothing (see below) as a solution to this problem.  

The downside of smoothing (adaptively or not) is that the power will be suppressed, or `smeared' out, on scales smaller than the kernel size. This is illustrated in Figure~\ref{fig:shot_noise} where we compare an un-smoothed power spectrum to one which is derived from an adapative SPH-smoothed map. To indicate the scale at which smoothing has a significant effect on the measurement (which we designate as a difference of 50\%), we plot a black, solid line.  However, we would like to stress that, although the power spectra are significantly affected on small scales, the comparison between the simulations and observations, which have both been smoothed in an equivalent way, is still valid even on small scales.

\subsubsection{Galaxy density}
Galaxies in a selected sample are smoothed in two-dimensional RA/DEC space with an adaptive smoothing kernel (SPH smoothing), the size of which is determined by the distance to $\mathrm{N_{sph}^{th}}$ nearest neighbour. (The same scheme is used in numerical simulations to derive 3D density estimates of particles.) We choose $\mathrm{N_{sph}=20}$ and $10$ for spectroscopic and photometric samples, respectively. (Through experimentation, we have found that these are approximately the minimum values that we can adopt for the two selections whilst retaining reliable estimates of the auto- and cross-spectra.)   The same values are used for simulated analogues. This allows for sparsely populated regions to be filled in with low density values, while dense regions are not over-smoothed so that the small-scale signal is preserved. SPH smoothing is described in more detail in Appendix~\ref{appendix:sph_smoothing}.  The smoothed density field is then projected onto a flat grid spanning the full extent of SDSS DR7 main survey footprint, i.e. $\mathrm{RA=[100,280]}$, $\mathrm{DEC=[-20,80]}$ degrees. The resolution of this grid is twice that of the \texttt{HEALPix} pixels, i.e. $0.858$ arcmin; this is done to ensure that all \texttt{HEALPix} pixels in the footprint are sampled and a contiguous footprint is obtained when we map the flat grid onto a \texttt{HEALPix} map.

The projection from the flat grid onto a \texttt{HEALPix} map is done by assigning pixel centre coordinates to the closest pixel centre in \texttt{HEALPix} via the inbuilt \verb'ang2pix' function. Square pixel values which are assigned to the same \texttt{HEALPix} are summed together.
On average, $\sim 4.6$ square pixels are assigned to one \texttt{HEALPix} pixel, however, $1.5\%$ of \texttt{HEALPix} pixels are singly-occupied. So coverage is far from uniform due to geometry. The total number of galaxies is conserved at all stages of pixelisation\footnote{All maps and masks will be made available for download at: \url{http://www.astro.ljmu.ac.uk/env_cor/}}.

\begin{figure*}
\includegraphics[width=0.98\columnwidth]{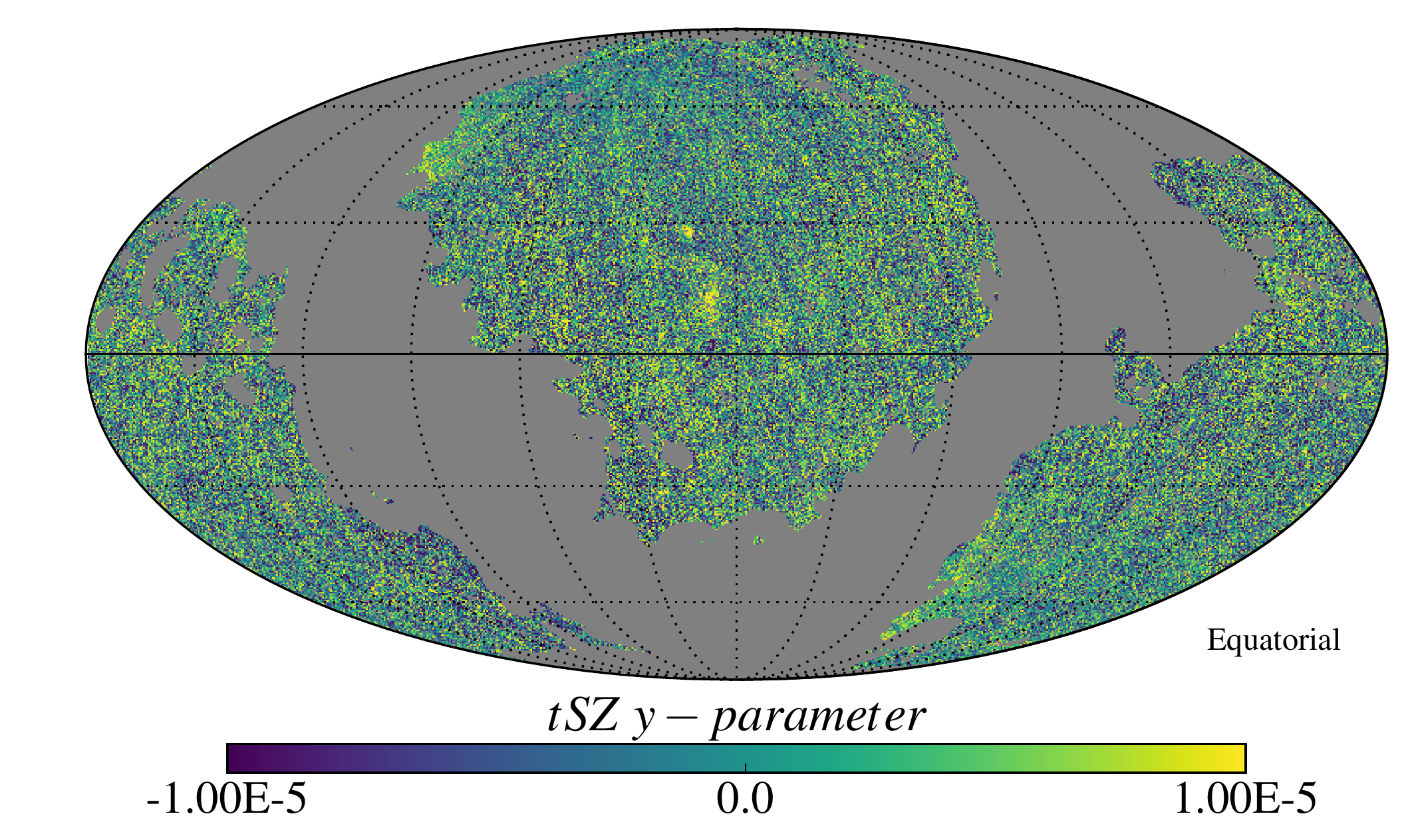} ($a$)
\includegraphics[width=0.98\columnwidth]{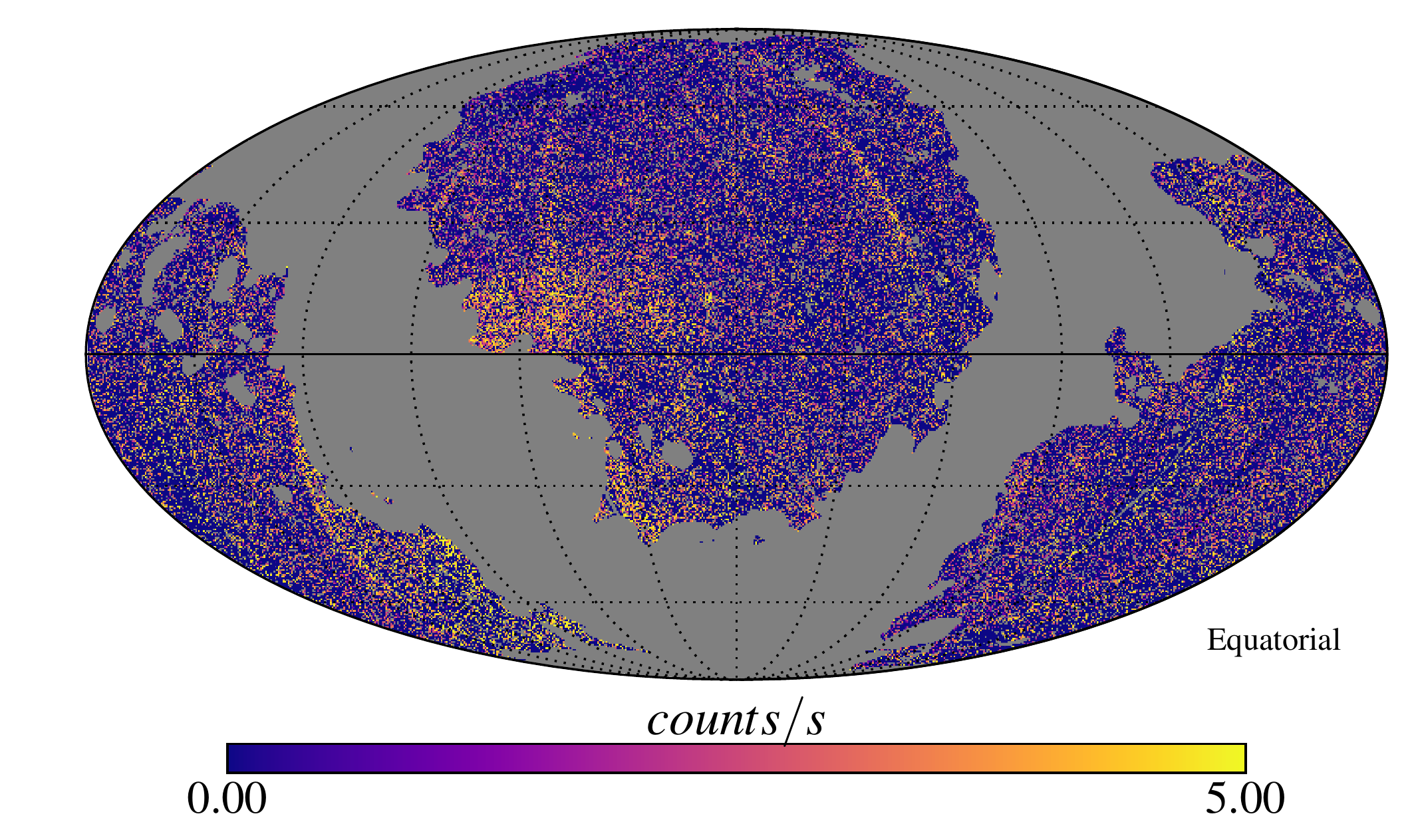} ($b$) \\
 ~ \\
\includegraphics[width=0.98\columnwidth]{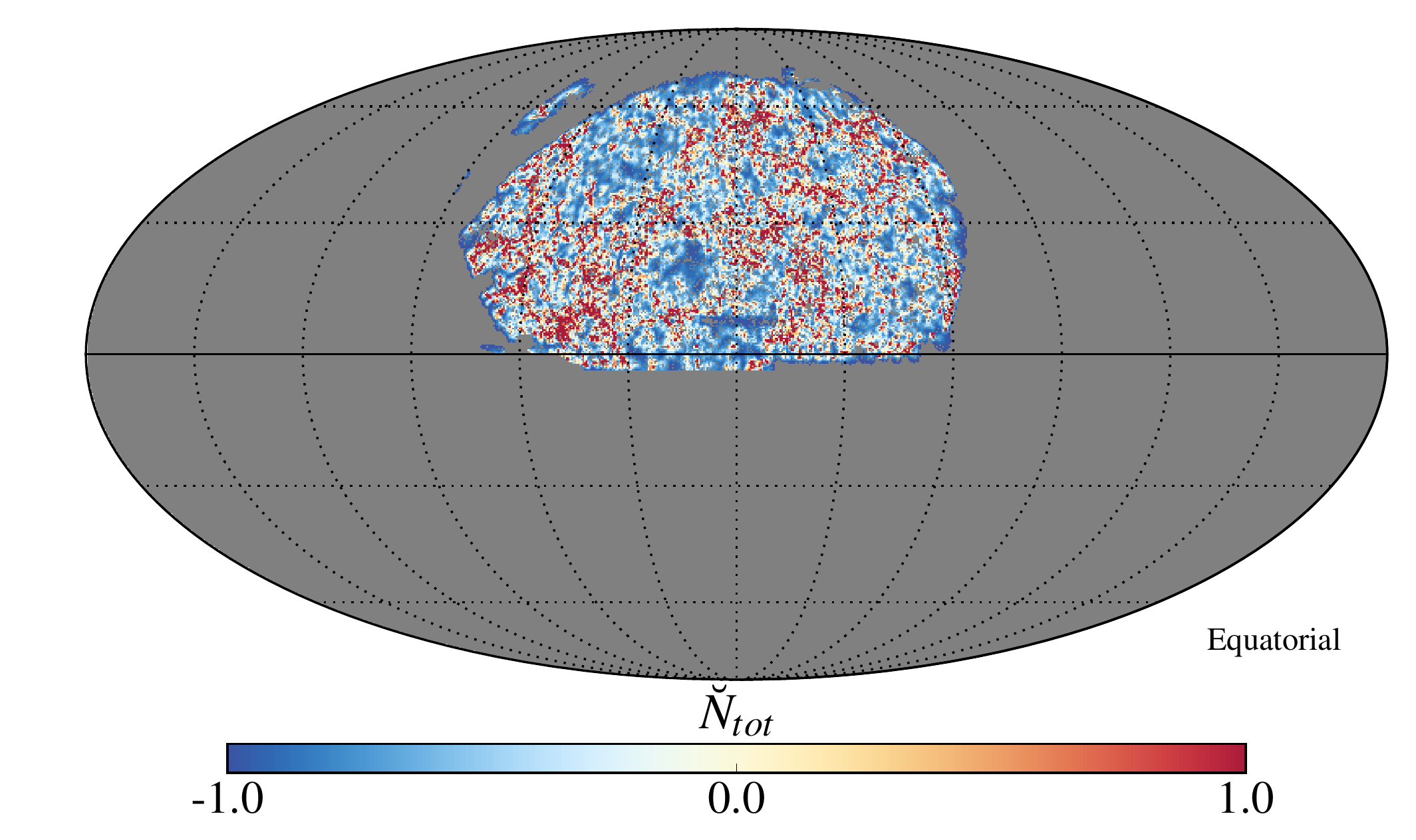} ($c$)
\includegraphics[width=0.98\columnwidth]{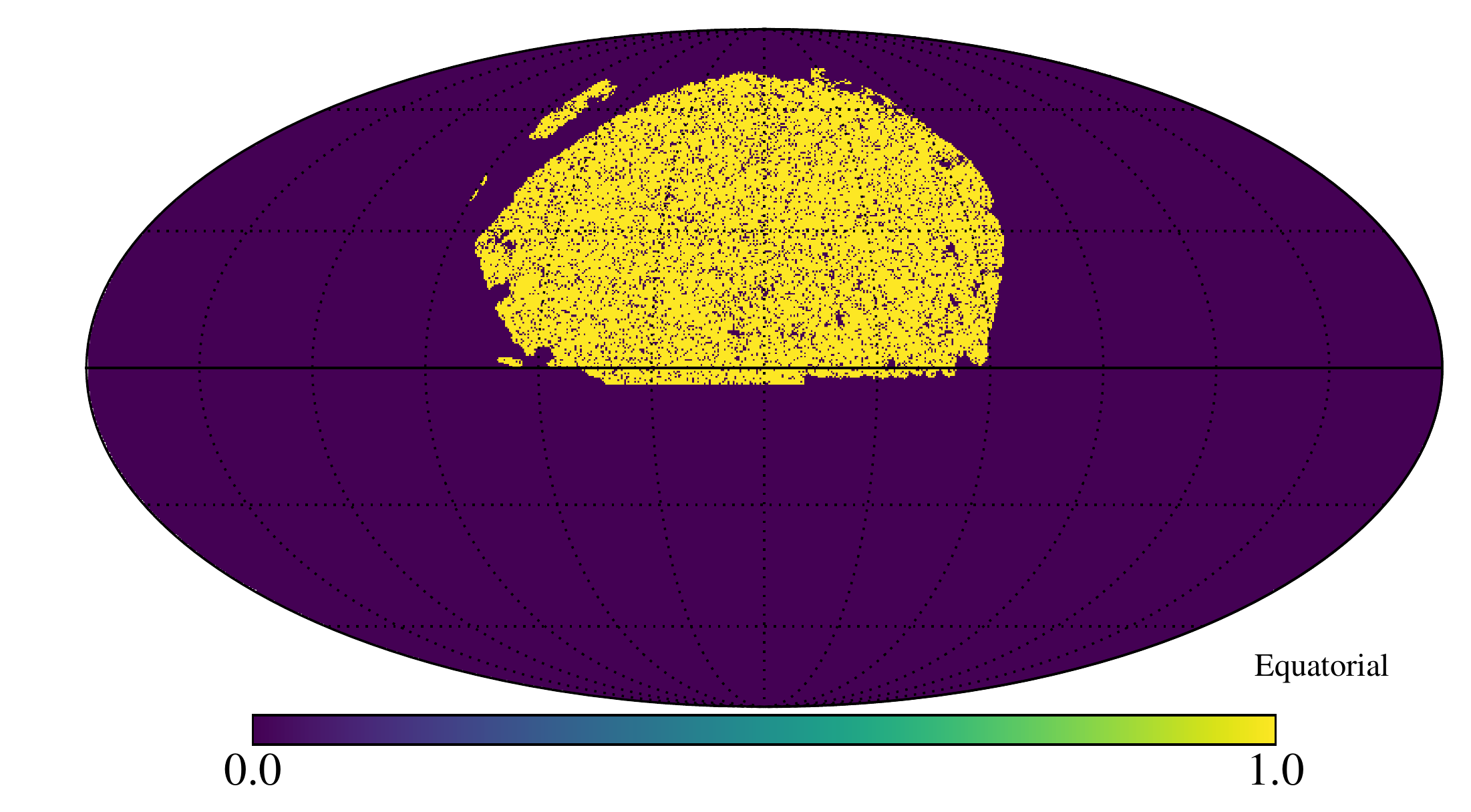} ($d$)
\caption{(a) A map of mean-subtracted thermal Sunyaev-Zel'dovich $y$ parameter as produced by the MILCA component separation algorithm from $Planck$ 2015 data.  Milky Way and point source masks have been applied (grey regions). The map has also been transformed into equatorial coordinates and rotated so that the region overlapping with SDSS footprint is in the centre. (b) A map of mean-subtracted X-ray flux in counts/s from RASS. Milky Way and point source masks have been applied (grey regions). The map has also been transformed into equatorial coordinates and rotated so that the region overlapping with SDSS footprint is in the centre. (c) A map of SPH-smoothed SDSS galaxies in the spectroscopic sample (galaxy overdensity), projected onto the \texttt{HEALPix} sphere. The final 7849 square degree footprint is as a result of a combination of the Milky Way, point source, and main SDSS footprint coverage. (d) Final mask resulting from the combination of Milky Way, point source, main SDSS footprint, and zero exposure by ROSAT masks. This mask gives a sky fraction of $0.16$ and is used in all correlations for respective galaxy samples in this paper.  These and other maps will be made publicly-available at \url{http://www.astro.ljmu.ac.uk/env_cor/}.}
\label{fig:maps}
\end{figure*}
\subsubsection{Quenched fraction}
As galaxies in the total and quenched samples are expected to have different clustering properties and, therefore, different SPH smoothing kernels if smoothed separately, it is necessary to take measures to keep the kernel consistent between maps. For this purpose, we construct a map of quenched flags that is smoothed simultaneously with the map for the total galaxy sample. Each quenched galaxy is assigned a binary flag ($1=\mathrm{quenched}$ or $0=\mathrm{star-forming}$) which forms the basis of our quenched fraction maps. Regions with high density of galaxies will be smoothed with a relatively small kernel, averaging the binary flags to a number between 0 and 1. A small kernel preserves the compact regions of highly-clustered, quenched galaxies (in contrast to a fixed-size Gaussian kernel) resulting in a high quenched fraction value, which is where most of the signal is expected to originate.

\subsection{Masking}

In addition to the regions of the sky not observed as part of SDSS (which is the main limiting survey in terms of area coverage in this study), two other masks are applied to the maps prior to computing power spectra.  First, we use the same Galactic mask\footnote{COM\_Mask\_Compton-SZMap\_2048\_R2.00.fits} as was used by the $Planck$ collaboration \citep{Planck_2014_tSZ}, specifically, the $\sim 40 \%$ sky Galaxy mask (M1), combined with the $Planck$ point-source mask (M5), yielding a sky fraction of $\sim 58 \%$.  We then combine this mask with the main SDSS footprint, which covers $\sim 19 \%$, giving a common area of $\sim 16 \%$ of the sky.  The $Planck$ galactic+point source mask, therefore, does not have a significant effect on the total available sky fraction.  

Finally, we also mask regions which were not observed by ROSAT as part of the RASS (i.e., those fields that have an exposure time value of zero). This masks a further $0.02 \%$ of the sky, which is negligible in terms of area but is necessary to avoid non-physical values.
The final mask used in this study can be seen in Figure \ref{fig:maps} (panel $d$).  This mask is consistently applied to all maps followed by mean-subtraction - to ensure it remains zero.  A map of galaxy overdensity for the SDSS spectroscopic sample is presented in panel ($c$) of the same figure.

\subsection{Thermal Sunyaev-Zel'dovich effect map}
The thermal Sunyaev-Zel'dovich (tSZ) effect \citep{Sunyaev_and_Zeldovich_1969} is a spectral distortion of the otherwise black-body CMB spectrum which is due to the inverse-Compton scattering of CMB photons by hot, free electrons (e.g., in the intracluster medium). The resulting change in intensity is directly proportional to the integrated line-of-sight electron gas pressure.  The (frequency-independent) amplitude of the tSZ effect is characterised by the dimensionless Compton $y$ parameter:
\begin{equation}
y = \frac{\mathrm{\sigma_T}}{\mathrm{m_e c^2}} \int \mathrm{P_e} \mathrm{dl},
\label{eq:tSZ}
\end{equation}
\noindent where $\mathrm{P_e \equiv n_e~k_BT_e}$ is the electron pressure (i.e., the product of the electron density and temperature); $\mathrm{\sigma_T}$, $\mathrm{m_e}$, and $\mathrm{c}$ are the conventional constants for Thomson scattering cross section, electron mass, and speed of light, respectively.

We use the publicly-available $Planck$ 2015\footnote{http://pla.esac.esa.int/pla/} MILCA \citep{Planck_sz_2016} tSZ effect maps and masks.  The data is stored in the \texttt{HEALPix} \citep{Gorski_Hivon_2011} format of $\mathrm{NSIDE}=2048$, so there were minimal adjustments made before cross-correlating.  In order to minimise radio continuum contamination from the Milky Way, a Galactic mask is used to mask $40\%$ of the sky, with an additional $\sim 2\%$ covered by the radio point-source mask, as mentioned previously. A masked version of the tSZ effect map is presented in Figure \ref{fig:maps} (panel $a$). 

\subsection{X-ray map}

The hot gas in and around galaxies and groups and clusters of galaxies emits radiation at X-ray wavelengths via thermal bremsstrahlung and recombination lines.  The observed X-ray surface brightness depends on the line-of-sight integrated electron density squared, as:
\begin{equation}
S_X = \frac{1}{4\pi(1+z)^4}\int n^2_e \Lambda(T_e, Z)dl,
\end{equation}
\noindent where $z$ is redshift, $n_e$ is the electron number density, and $\Lambda(T_e, Z)$ is the emissivity (or cooling function) in the relevant energy band, which only weakly depends on the temperature and metallicity (Z) of the gas for the energy range of $0.4 - 2.4~\mathrm{keV}$ \citep{Adam_etal_2017} sampled by ROSAT.  Note how rapidly the surface brightness drops off with redshift ($S_X \propto (1+z)^{-4}$), making individual system analysis prohibitive for anything other than the most nearby objects and/or the most massive clusters.

We point out the differing dependencies of the X-ray emission and tSZ effect on the properties of the hot gas (density and temperature).  In principle, examination of cross-correlations between galaxy properties and these two observables should therefore help to break degeneracies in environmental processes (e.g., ram pressure stripping does not depend on the gas temperature).

Full-sky X-ray observations are derived from the ROSAT All-Sky Survey (RASS)\footnote{https://heasarc.gsfc.nasa.gov/docs/rosat/rass.html} (see \citealt{Voges_1993} and \citealt{Voges_etal_1999} for survey description).  The survey was completed using the Position Sensitive Proportional Counter (PSPC) instrument aboard the ROSAT satellite in `scanning' mode. The original RASS data is organised into $6.4\times 6.4$ degree fields of the sky, which have been assembled into a full-sky map\footnote{http://www.xray.mpe.mpg.de/cgi-bin/rosat/rosat-survey} in world coordinate system (WCS).  This full-sky map has been conveniently projected into \texttt{HEALPix} format by the Centre d'Analyse de Donn\'ees Etendues (CADE)\footnote{http://cade.irap.omp.eu/dokuwiki/doku.php?id=rass}. The process is based on the drizzling library \texttt{Drizzlib}\footnote{http://cade.irap.omp.eu/dokuwiki/doku.php?id=software} and the technical aspects of transforming from the WCS to \texttt{HEALPix} format are described in Appendix A of \citet{Paradis_etal_2012}. This method guarantees photometric accuracy of the transformation with minimal data loss during the conversion from a local WCS FITS map to HEALPix format. We use the $0.4 - 2.4~\mathrm{keV}$ count map (RASS\_IM2\_1\_2048.fits) together with the exposure time map (RASS\_EXP\_1\_2048.fits). To obtain a map of count rate, we simply divide counts by exposure and multiply by 5.24559 (see CADE website) to obtain the photon rate per \texttt{HEALPix} pixel, masking regions where exposure is equal to zero.
As already noted, in order to make consistent comparisons between the two environment maps (tSZ effect and X-ray), the zero-exposure X-ray mask is combined with the Galaxy and point-source masks from $Planck$. A map of RASS with the Galaxy, point source, and zero-exposure masks applied can be seen in Figure \ref{fig:maps} (panel $b$).

\subsection{Simulations}
In order to decompose our detected signals and relate them to three-dimensional processes acting on single cluster scales, we utilise two different cosmological hydrodynamic simulations. Comparison to simulations is also beneficial in guiding our observational analysis, since the simulation-based maps do not have masks or incomplete coverage and allow us to experiment with the effects of smoothing and noise.

\subsubsection{EAGLE}
EAGLE (Evolution and Assembly of GaLaxies and their Environments,
\citealt{Schaye_etal_2015,Crain_etal_2015}) is a suite of cosmological,
hydrodynamic simulations designed to study the formation and evolution of galaxies at high resolution in a moderate-size box.  Its standard (Ref-L100N1504) scientific run consists of a $\mathrm{L=}100~\mathrm{comoving\ Mpc\ (cMpc)}$ (on a side) box, with $\mathrm{N=}1504^3$ collisionless dark matter particles and an equal number of baryonic particles. The simulations were carried out with a modified version of the Lagrangian Tree-SPH code \texttt{GADGET3} (last described by \citealt{Springel_2005}).
EAGLE adopts a $Planck$ 2013 \citep{Planck_cmb_2014} $\mathrm{\Lambda CDM}$ cosmology.  

The EAGLE simulations include a number of subgrid treatments of processes that cannot be directly resolved in the simulations, including metal-dependent radiative cooling, star formation, stellar evolution and mass-loss, BH formation and growth, and stellar and AGN feedback. The efficiency of stellar feedback was calibrated to approximately reproduce the local galaxy stellar mass function and the sizes of galaxy discs, while the efficiency of AGN feedback was calibrated to reproduce the present-day scaling relation between the stellar mass of galaxies and that of their central BH (for further details see \citealt{Crain_etal_2015}).  The feedback was not calibrated to reproduce the hot gas properties of galaxies or groups and clusters.  It was recently shown in \citet{Davies_etal_2019} that simulations tend to predict X-ray luminosities that are somewhat too high compared to those observed for optically-selected samples of galaxies, whilst the simulations reproduce the large-scale tSZ effect flux well.  We return to this point later.

\subsubsection{BAHAMAS}
The BAHAMAS (BAryons and Haloes of MAssive Systems, \citealt{McCarthy_etal_2017,McCarthy_etal_2018}) project is another set of cosmological, hydrodynamic simulations based on \texttt{GADGET3}, however its focus is on large-scale structure cosmology.  As such, the simulations consist of much larger volumes but at significantly lower resolution than EAGLE.  Specifically, the simulations primarily consist of $400~\mathrm{cMpc/h}$ periodic boxes containing $2 \times 1024^3$ particles (dark matter and baryonic, in equal numbers).  Here we use the run that adopts a WMAP9 \citep{Hinshaw_etal_2009} cosmology with massless neutrinos.

BAHAMAS includes subgrid treatments for all of the same processes mentioned above for EAGLE, though with somewhat different parametrisations.  The subgrid models were developed as part of the OWLS project \citep{Schaye_etal_2010}.  The parameters governing the efficiencies of AGN and stellar feedback were adjusted so that the simulations reproduce the observed galaxy stellar mass function for $\mathrm{M_*} \geq 10^{10}~\mathrm{M_{\odot}}$ and baryon content of groups and clusters, as dictated by the gas mass fraction--halo mass relation from high-resolution X-ray observations.  As shown in \citet{McCarthy_etal_2017}, the simulations match the local X-ray and tSZ effect scaling relations of galaxies and groups and clusters.

\subsubsection{Light cones and simulated map making}
\label{light_cones_and_sim_maps}

To make like-with-like comparisons between the observations and simulations, equivalent maps of X-ray, tSZ effect, and galaxy properties from the simulation data are required.
This is achieved by constructing light cones, extending from a point (the simulated observer) in a single direction of the simulation. A single simulation box does not have the required depth to accommodate a light cone out to $\mathrm{z} = 0.15$\footnote{This is the maximum redshift of the galaxy catalogue, resulting in the maximum cross-correlation signal possible. Any additional depth in X-ray/tSZ maps (as is the case with observations) only results in additional noise on the cross-power spectra, this was tested by constructing light cones out to $\mathrm{z}=3$ and cross-correlating with a galaxy catalogue truncated to a lower $\mathrm{z}$ - noise increased with increasing discrepancy but the mean cross-power spectra did not change.}, so several simulation boxes have to be stacked in a line.  Ideally, different sets of simulations would be used for each box constituting a light cone, however this would be too computationally expensive.  As a compromise, randomly translated/rotated/reflected snapshots of the same simulation are used. This minimises box-to-box correlations, so that they appear to be independent realisations of the simulated universe. Further detail on light cone construction can be found in \citet{McCarthy_etal_2018} (see also \citealt{DaSilva_etal_2000}). 

Ten different light cones are constructed for each simulation, representing 10 different lines of sight, allowing for an estimate of cosmic variance.  Due to BAHAMAS having a simulation volume that is significantly larger than EAGLE, the field of view per light cone is $25~\mathrm{degrees}$ on a side for the former, and $10~\mathrm{degrees}$ for the latter.  Cone-to-cone variance is, therefore, greater for EAGLE than BAHAMAS.

The desired quantities within a light cone now need to be projected and mapped onto 2D pixels. We follow \citet{McCarthy_etal_2018} when making the simulated tSZ Compton $y$ maps. The parameter is computed directly from the properties of gas particles, first performing the integral in equation \ref{eq:tSZ} and dividing that by the area of a pixel at the angular diameter distance of the particle (see \citealt{McCarthy_etal_2018} for details).

To compute X-ray maps, we first compute high resolution (dE = 2 eV) synthetic X-ray spectra spanning the range 0.4-40.0 keV for each hot gas particle using the Astrophysical Plasma Emission Code \citep[APEC;][]{Smith_etal_2001} with updated atomic data and calculations from the AtomDB v2.0.2 \citep{Foster_etal_2012}. The spectrum of each gas particle is computed using the particle's density, temperature, and full abundance information.  (Note that we exclude cold gas below $10^5$ K which contributes negligibly to the total X-ray emission.)  The spectra are appropriately redshifted using the redshift of the gas in the light cone and converted from intrinsic luminosity units into observed photon flux, taking account of cosmological dimming.  As the RASS maps are provided in counts/s in the $0.4-2.4$ keV, it is also necessary to fold in the instrumental response of the ROSAT PSPC instrument.  To achieve this, we convolve the synthetic spectra with the effective area vs.~energy function for the PSPC instrument, obtained from the WebPIMMS Count Rate Simulator\footnote{https://heasarc.gsfc.nasa.gov/cgi-bin/Tools/w3pimms/w3pimms.pl}.  This provides an estimate of the observed count rate in the 0.4-2.4 keV band as would be observed with ROSAT.  We sum the contribution of each hot gas particle along the line of sight to the observed count rate, as was done for the tSZ effect.

When constructing the tSZ effect and X-ray maps, the initial native pixel size adopted was 10 arcseconds, which we rebinned to size of $1.7~\mathrm{arcmin}$ in order to match the mean resolution of \texttt{HEALPix} pixels of tSZ effect and X-ray maps used in this study. The mapping of particles to a grid is done using a simple `nearest grid point' interpolation method.  Finally, the maps are smoothed with Gaussian beams of 10 arcmin for the tSZ effect maps for consistency with the $Planck$ maps, and $1.8$ arcmin for the X-ray maps for consistency with the PSF of the ROSAT PSPC instrument. In the case of \emph{Planck} tSZ map, there is a dominant noise component contributing both positive and negative values in the map. The physical signal only contributes positive values. We, therefore, fit a Gaussian to the negative side of the pixel distribution, mirror it to the positive side, then sample this function to draw noise values for each pixel in the simulated maps.  Maps are mean-subtracted after noise addition to ensure a mean of zero, as for the \emph{Planck} map.

In terms of galaxy catalogue-based maps, we select all galaxies in the light cones which have a mass exceeding $10^{10} \rm{M_\odot}$ and lying within either $z<0.06$ (spectroscopic) or $z<0.15$ (photometric).  Galaxies are defined to be either star-forming or quenched on the basis of their sSFR within a 30 kpc aperture, using the same threshold ($10^{-11}~\rm{yr^{-1}}$) as employed for the (spectroscopic) observations.  Galaxies and their quenched flags are deposited into maps using the same SPH smoothing algorithm employed on the SDSS data. The value for $\mathrm{N_{sph}}$ (number of smoothing neighbours) was chosen such that a contiguous field is obtained with minimal smoothing. In this case, we use $\mathrm{N_{sph}}=20$ and $10$ for spectroscopic and photometric samples, respectively, as was done for the observational data.

As we use overdensities/quenched fraction excess in the cross-correlations, it is necessary to compute the mean which we then use as a denominator in our calculations. This makes the normalisation of power spectra sensitive to the value of the mean. Given the degree of cosmic variance in the simulations, especially EAGLE, using the mean value of each light cone results in a substantially different normalisation of the power spectra. Since our goal is to use simulations in the interpretation of observed signals, it is desirable to have the power spectra as close as possible between simulations and observations. For this reason, we adopt the mean values of $\mathrm{N_{tot}}$ and $\mathrm{f_{q}}$ from SDSS maps (which are not limited by cosmic variance) and use them to compute excesses in the simulations.

\section{Auto- and Cross-power spectra estimation}
\label{section:cross-correlation}
\subsection{Formalism}
We employ a two-point statistic to make a quantifiable measure of the correlation between two maps, each of which contains scalar quantities, using a quadratic estimator \citep{Chiang_etal_2011}, i.e.:
\begin{equation}
C^{jj'}_l = \frac{1}{2\ell + 1}\sum^{\ell}_{m=-\ell}  j^*_{lm} j'_{lm} \equiv \lvert j_{lm} \rvert \lvert j'_{lm} \rvert \cos \Delta \phi_{lm},
\label{eq:cross-spectrum}
\end{equation}
\noindent where $j$ and $j'$ are the two maps being considered, $C^{jj'}_l$ are the cross-power spectrum coefficients in multipole, $l$, space. $\Delta \phi_{lm}$ is the phase between $j$ and $j'$, and takes values in the interval $[0,2\pi]$; in a case where $j$ and $j'$ are the same signal, $\cos \Delta \phi_{lm}$ returns $1$, and $0$ if signals are uncorrelated. The effect is such that $\langle \lvert j_{lm} \rvert \lvert j'_{lm} \rvert cos \Delta \phi_{lm} \rangle = 0$ in the case of spatially uncorrelated maps (where angle brackets indicate ensemble averages), otherwise it becomes a quantifiable measure of correlation between them.  Note that if the two maps, $j$ and $j'$, are identical, we obtain an estimate of the auto-power spectrum.  If the maps differ, we estimate the cross-spectrum.

Values for $j$ and $j'$ could be computed directly from the maps if they were available for the entire sky, however, only some surveys have observed the entire sky and even then there are foreground objects/contaminants which need to be masked. Masking has the effect of mode mixing, whereby eqn.~\ref{eq:cross-spectrum} becomes an estimate of the biased \emph{pseudo-}power spectrum and needs to be corrected for.
The incomplete sky window-function can be represented as a position-dependent weighting with its own power spectrum:
\begin{equation}
W_l = \frac{1}{2\ell + 1}\sum^{\ell}_{m=-\ell} \lvert w_{lm} \rvert^2,
\end{equation}
where $w_{lm}$ are the spherical harmonic coefficients of the window function that is convolved with the underlying map of interest \citep{Hivon_etal_2002}. The spherical harmonic coefficients, in this case, take the following form:
\begin{equation}
\tilde{j}_{lm} = \int \Delta J(\hat{\mathbf{n}})W(\hat{\mathbf{n}}) Y^*_{lm}(\hat{\mathbf{n}})d\hat{\mathbf{n}} \approx \Omega_p \sum_p J(p)W(p) Y^*_{p},
\end{equation}
where $J(\hat{\mathbf{n}})$ is the scalar quantity captured in the map (and $\Delta J(\hat{\mathbf{n}})$ is its fluctuation from the mean), $W(\hat{\mathbf{n}})$ is the window function, and $Y^*_{lm}(\hat{\mathbf{n}})$ represents the spherical harmonics. In this equation we also write the expression for quantised maps where $p$ represents a pixel, $\Omega_p$ is the pixel area, and the sum is over all data pixels in the map.
The challenge is then to correct for the effects of partial-sky observations.  Fortunately, there are existing publicly-available algorithms to do so.


\verb'NaMASTER'\footnote{https://github.com/LSSTDESC/NaMaster} \citep{Hivon_etal_2002, Alonso_etal_2018} is an algorithm based on the direct spherical harmonic transform of the input maps. It operates entirely in the spherical harmonic space and, among many other functions, performs mask-correction by inverting the mode-mixing matrix relating the pseudo-power spectrum with the full-sky power spectrum estimator:

\begin{equation}
\langle \tilde{C}_l \rangle= \sum_{l'} M_{ll'} \langle C_l \rangle.
\end{equation}

In order to reduce windowing effects when performing the inversion, it is necessary to perform mask apodization prior to the computation. \verb'NaMASTER' has multiple modes of apodization built in. For our combined mask, it was found that the mode `C1' with $\mathrm{apodization\ parameter}\ \theta_* = 0.04$ is the most optimal in recovering the full-sky power spectrum estimate. This mode involves multiplying all pixels by a factor $f$ given by:
\begin{equation}
f(x)= 
\begin{cases}
    x - \mathrm{sin}(2\mathrm{\pi} x)/(2\mathrm{\pi}),& \text{if } x < 1\\
    1,              & \text{otherwise}
\end{cases}
\end{equation}
where $x \propto \sqrt{(1 - cos \theta) (1 - cos \theta_*)}$, and $\theta$ is the angular separation between the pixel and its closest masked neighbour.

Apodization was calibrated by making a simulated map from a known power spectrum, applying our mask, and demanding that the power spectrum be recovered within $1 \%$ error. For this purpose we made use of the best-fit $\mathrm{\Lambda CDM}$ CMB TT power spectrum provided by the $Planck$ collaboration\footnote{http://pla.esac.esa.int/pla/\#cosmology}, as it has many of the same statistical properties as the tSZ effect map. A full-sky map of $\mathrm{N_{SIDE} = 2048}$ was generated and fed into the power spectrum estimator prior to computing all of the power spectra presented in this paper.

In terms of cross-correlations between hot gas (X-ray, tSZ) and galaxies, the methodology we have adopted, which is projecting galaxy surveys onto a HEALPix map for estimates of the Fourier-based cross-spectra with tSZ and X-ray data already in the HEALPix format, is fast becoming the standard practice (e.g., \citealt{Makiya2018,Koukou2020,Pandey2020}).  However, given that the galaxies are treated as discrete objects, in principle one does not need to project the galaxies onto a map to analyse their clustering signal or cross-correlations with other signals.  We have opted for a map-based approach not because there are obvious flaws with an object-based analysis, but mainly for convenience.  Specifically, our map-based approach was motivated by: i) the available hot gas data we use ($Planck$ Compton $y$ and RASS X-ray counts) are in map form and, in the case of the $Planck$ $y$ map it has been smoothed to a fixed resolution of 10 arcmin; ii) from the point of view of galaxies, we are mainly interested not in individual properties but in ensemble quantities such as quenched fraction, so some spatial averaging is required (which is easily achieved within a map framework); iii) there are a wide variety of existing well-tested, publicly-available tools for  analysing HEALPix maps (such as \verb'NaMASTER' and PolSpice); and iv) we can take advantage of our own software for projecting large cosmological hydrodynamical simulations onto maps (e.g., \citealt{McCarthy_etal_2014,McCarthy_etal_2018}), thus allowing for a relatively straightforward comparison with the available data.

We expect that, if handled correctly, object-based and map-based analyses should converge on spatial scales above the pixel scale, which is where we limit our analysis to in the present study.  It would be interesting to directly compare object-based and map-based approaches, but we leave this for future work.

\subsection{Null-tests and error estimation}
\label{section:null_test}
Given the noisy nature of the observational maps used here, there is always a possibility of obtaining a non-zero cross-power spectrum when there is no physical correlation. In order to ensure that our signals are real and not just driven by noise, we perform random rotations of one map relative to the other and compute power spectra for these combinations.  Since no physical correlation is expected when the maps are rotated with respect to each other, any correlation that does persist is a result of noise.  When rotating the maps, we ensured that galaxy/point source masks were fixed where necessary and a new combined mask was made prior to each computation. Each null power spectrum was inspected to check that it is consistent with zero over the entire $\mathrm{\ell}$ range when binned in the same way as the signal power spectra.
We indeed observe that each null power spectrum oscillates around zero, rarely having the same sign in several consecutive $\mathrm{\ell}$ bins.  We estimate the $1 \sigma$ regions from $100$ rotations for every correlation we compute and only plot the upper part of the region as all of our auto-/cross-power spectra are positive.  The distance away from the null-test $1 \sigma$ region provides a visual demonstration of the significance of any detection.  The turquoise regions in Figs. \ref{fig:panel_nex_fq} and \ref{fig:panel_tsz_x} (below) represent the $1 \sigma$ uncertainties in the auto- and cross-spectra as derived from the null tests.

We analytically estimate the statistical error bars on the observed and auto- and cross-spectra following the formalism of \citet{Tristram_etal_2005} (see also \citealt{Hill_etal_2014,Hurier2015}). For auto-spectra ($\mathrm{\breve{N}_{tot}}$ and $\mathrm{\breve{f}_{q}}$) this involves computing (see eqns.~29-32 in \citealt{Tristram_etal_2005}):

\begin{equation}
(\Delta C^{j j}_{\ell})^2 = \frac{1}{f_{\rm sky}} \frac{2}{(2\ell +1)\Delta \ell} (C^{j j}_{\ell})^2,
\end{equation}

\noindent where $f_{\rm sky}$ is the unmasked fractional area of the sky, $\Delta \ell$ is the width of a multipole bin centred on $\ell$, and $C^{j j}_{\ell}$ is the power spectrum estimate.

Similarly, for the cross-spectra, the statistical errors are estimated using:

\begin{equation}
(\Delta C^{jj'}_{\ell})^2 = \frac{1}{f_{\rm sky}} \frac{2}{(2\ell +1)\Delta \ell} \left( C^{jj}_{\ell} C^{j'j'}_{\ell} + (C^{jj'}_{\ell})^2\right),
\end{equation}

\noindent where $C^{jj}$ is auto-spectrum of the first map, $C^{j'j'}$ auto-spectrum of the second, and $C^{jj'}$ is the cross-power spectrum.

We note that the formalism of \citet{Tristram_etal_2005} was originally designed with the analysis of CMB maps in mind (including tSZ maps), as opposed to galaxy surveys.  Therefore, as a check, we have also performed a `brute force' estimation of the uncertainties by performing 100 random realisations of the galaxy density field under the assumption of Poisson statistics.  That is, for each pixel we draw from a Poisson distribution whose mean is equal to the number of galaxies in that pixel.  We generate 100 randomised realizations of the original SDSS galaxy density field in this way.  
For the galaxy density (auto-)power spectrum in particular, we find that the \citet{Tristram_etal_2005} and Poisson resampling uncertainties agree to typically better than a factor of 2 over the full multipole range, with both being $\approx0.01$ of typical power spectrum measurement (i.e., the galaxy power spectrum is very strongly detected).  While the two methods of calculating uncertainties do not yield identical results (for undetermined reasons), none of the qualitative results or conclusions in this study are affected by our choice of error estimation technique.  For specificity, we show the uncertainties calculated using the widely-employed Tristram et al.~formalism throughout.

\section{Results}
\label{section:results}
\begin{figure*}
\includegraphics[width=\textwidth]{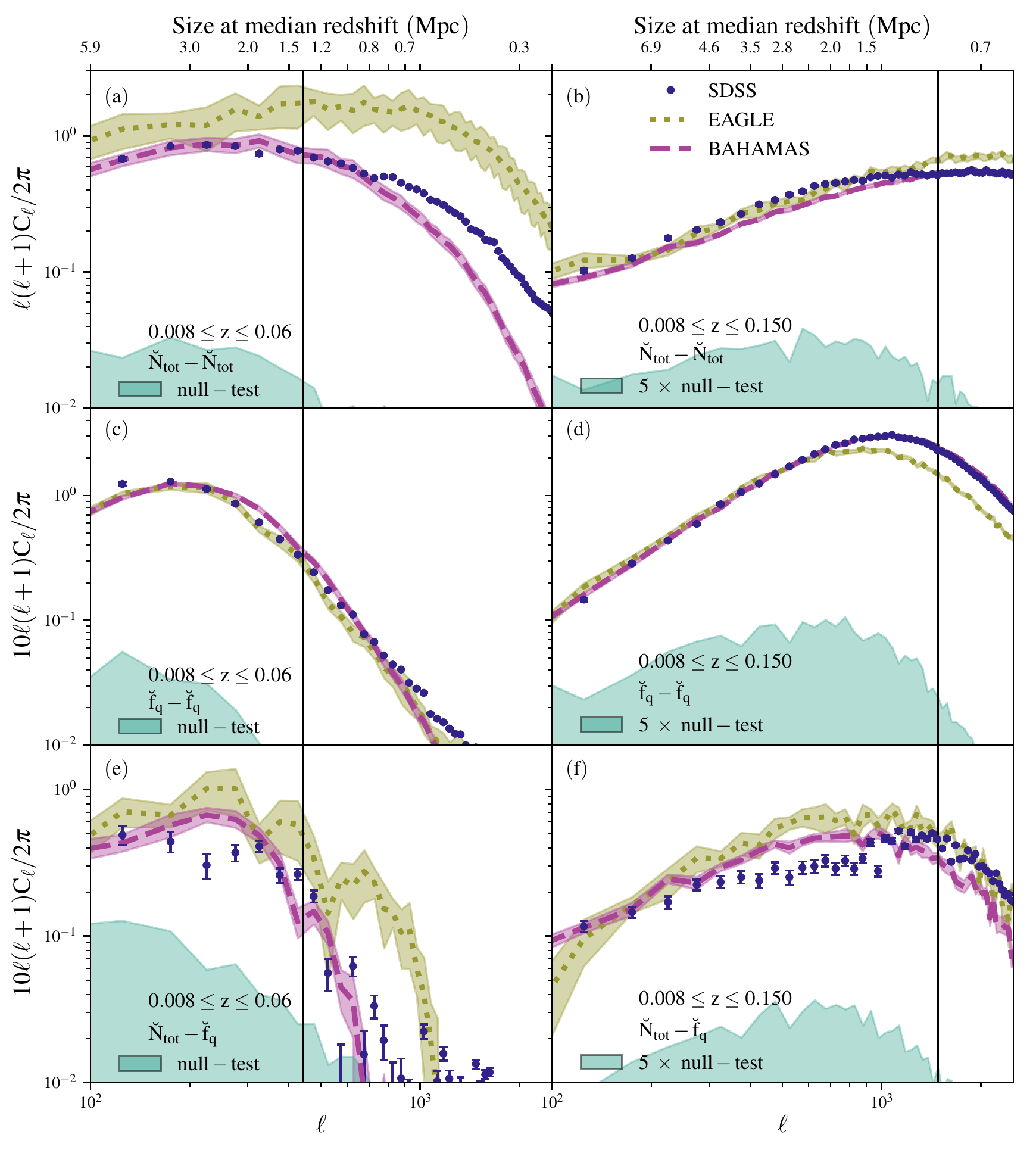}
\caption{Auto and cross-power spectra involving the spectroscopic (\emph{left}) and photometric (\emph{right}) SDSS galaxy samples. Navy data points with error bars show the measured signal from the SDSS galaxy samples; pink, dashed curves and yellow, dotted curves show the predictions from BAHAMAS and EAGLE, respectively.  Shaded regions indicate the $1\sigma$ confidence interval on the mean value of 10 light cones.  Turquoise, shaded regions indicate the $1\sigma$ level of the null-signal (see text).  Vertical, black lines indicate the scale at which SPH-smoothed power spectra deviate from their discrete counterparts by 50\% (see Appendix \ref{appendix:sph_smoothing}).  The top x-axis indicates the approximate physical scale (transverse) assuming the median redshift of each sample. Top row (panels~($a$) and ($b$)) shows the total, shot noise-subtracted, galaxy overdensity power spectra; Middle row (panels~($c$) and ($d$)) shows the quenched fraction excess power spectrum; Bottom row (panels~($e$) and ($f$)) shows the quenched fraction--galaxy overdensity cross-spectra.  Strong signals are detected in each case and are qualitatively consistent with the simulation predictions.  In detail, the simulations reproduce the galaxy overdensity power spectrum and quenched fraction excess power spectrum remarkably well (particularly BAHAMAS), but both simulations predicted somewhat larger than observed amplitudes for the quenched fraction--overdensity cross-spectrum.}
\label{fig:panel_nex_fq}
\end{figure*}

We present our results in the form of panel plots in Figures~\ref{fig:panel_nex_fq} and \ref{fig:panel_tsz_x}, comparing the spectroscopic and photometric estimates of various auto- and cross-correlations side-by-side. This highlights the similarities and differences between the two samples as well as making it easier to spot changes in physical scale for different quantities. 

In all cases, we restrict the multipole range between $100 < \ell < 2500$, which approximately corresponds to angular scales of $4.32 < \theta < 108$ arcmin\footnote{For reference, our pixel size is $1.716$ arcmin and SDSS fibre angular resolution is $3$ arcsec.}. On scales below $\ell \sim 100$ the observations become sparsely sampled and noisy, whereas the simulations reach their field of view limit leading to an abrupt truncation of the signal.
At high multipoles, the observed and simulated power spectra are affected by the SPH smoothing kernel applied to the galaxy distribution and a beam present in the tSZ effect or X-ray observations, when conputing cross-spectra invovling those quantities.  By $\ell \sim 2500$ these limiting factors are fully in effect and all power spectra smoothly decline towards zero with increasing $\ell$. There is little information to be gained from examining such small scales, hence our limit.

We begin by plotting galaxy and quenched fraction auto-power spectra, followed by their cross-spectrum. 
Next, we introduce measures of gas environment by computing the tSZ effect/X-ray cross-spectra with galaxy overdensity and quenched fraction. These measure the connection between galaxy overdensity and hot gas pressure and density.  The scales over which these quantities correlate indicates the angular scales over which the interplay between them acts. To further aid in this interpretation, we plot an approximate physical scale for these angular scales at the median redshift of the galaxy sample. 

The cross-correlation of quenched fraction (as opposed to galaxy overdensity) with hot gas properties takes this one step further, by examining the star-forming state of galaxies. A statistically significant signal here would be the first time that a connection between hot gas properties and the quenched state of galaxies is measured directly without first selecting regions of the sky known to contain groups and clusters.  This is important, as it is the local physical conditions that characterise the environment and not whether one has labelled that there is a group/cluster present. 

\subsection{Galaxy overdensity power spectrum}

Panel ($a$) in Figure~\ref{fig:panel_nex_fq} shows the galaxy overdensity (i.e., auto-$\mathrm{\breve{N}_{tot}}$) power spectrum for the SDSS spectroscopic sample (navy points with error bars).  Panel ($b$) shows the same quantity for the larger photometric sample.  
Note that the galaxy overdensity power spectrum is just the Fourier transform of the perhaps more familiar configuration-space (projected) two-point correlation function (2PCF).   All power spectra in panels ($a$) and ($b$) have been shot noise-subtracted.  (Please refer to Appendix~\ref{appendix:sph_smoothing} for a discussion of shot noise estimation.)  Also shown are the 1-sigma contours (turquoise shaded region) derived from the null tests (see Section \ref{section:null_test}) which give an additional estimate of the observational uncertainties due to noise, as are the predictions of the EAGLE and BAHAMAS simulations, for which we adopt the same selection criteria as in the observations.  All power spectra (observed or simulated) are suppressed on small scales due to SPH smoothing.  The scale at which this effect is $50\%$ or more is indicated by a vertical solid, black line.  Please refer to Appendix~\ref{appendix:sph_smoothing} for more discussion of the effects of smoothing.  Note that because smoothing is applied consistently to the observational and simulated maps, comparisons between the observations and simulations on scales smaller than the smoothing scale are still meaningful.  

As expected, strong signals are detected on all scales that we sample, with the $1\sigma$ error bars being generally smaller than the data points and the null-test $1\sigma$ limit being at least an order of magnitude lower than the signal (and considerably larger than this for the photometric selection) on all scales.  The observed power spectrum reaches a plateau at $\ell \sim 300~(1000)$ for the spectroscopic (photometric) selections and declines thereafter due to smoothing effects.  Note that differences are expected between the observed power spectra of the spectroscopic and photometric samples just on the basis that these are {\it angular} correlation functions and that the two samples have differing mean redshifts.  A secondary effect is that the angular scale where the effects of smoothing become pronounced is reduced for the deeper photometric selection.  This is just because the number of galaxies per pixel is increased and therefore the angular scale enclosing a fixed number of neighbours is decreased.

Both EAGLE and BAHAMAS reproduce the clustering of galaxies seen in the deeper photometric sample remarkably well.  Qualitative agreement is also found for the spectroscopic sample comparison (i.e. similar shape and amplitude as the observed sample), although clear quantitative discrepancies can be seen between the observations and the two simulations.  We attribute these differences to the larger degree of cosmic variance present in the simulations for the shallower (smaller volume) spectroscopic selection.  (Note that, as the SDSS footprint is of {\it much} larger area than either the EAGLE or BAHAMAS light cones, the cosmic variance errors for the observed power spectra are negligibly small.) 

The top panels of Figure~\ref{fig:panel_nex_fq} establish that galaxies in the simulations cluster in approximately the same way as in the real Universe.  Note that this is only expected to be true if galaxies trace the correct haloes in the simulations (i.e., they have the correct stellar mass--halo mass relation, so that the selected galaxies have the correct bias with respect to clustering of matter) and the adopted cosmology is also broadly correct (so that the simulations have the correct matter clustering).  The agreement of the predictions of the simulations with the observed galaxy overdensity power spectrum on small scales in particular may at first seem surprising, given the relatively large spread in predictions from hydrodynamical simulations for quantities like the total matter power spectrum at fixed cosmology (e.g., \citealt{Chisari2019,vanDaalen2019}).  However, it has been shown previously that many of the clustering statistics of galaxies can be reproduced relatively well by simple abundance matching techniques (see \citealt{Conroy2009} and references therein).  As both EAGLE and BAHAMAS have been calibrated to reproduce the observed galaxy stellar mass function, the resulting stellar mass--halo mass relations from these simulations agree well with abundance matching methods (see \citealt{Schaye_etal_2015,McCarthy_etal_2017}).  Consequently, the simulations should also reproduce the clustering statistics at least as well as abundance matching methods.    

Since the spatial distribution of galaxies is correct, we can go further and ask whether the galaxies in the simulations respond to the environment in the same way as real galaxies.

\subsection{Quenched fraction power spectrum}

In panels ($c$) and ($d$) of Figure~\ref{fig:panel_nex_fq} we show the clustering of the quenched fraction (i.e., auto-$\mathrm{\breve{f}_{q}}$).   Note that this is a measure of how quenched fraction itself clusters, independently of galaxy density. (Only the SPH smoothing kernel is common between maps of quenched fraction and overdensity.)  Having said that, from previous studies the quenched fraction and galaxy density are known to be correlated quantities.  It is therefore reasonable to expect that a similar correlation signal is observed here as in the top panels of Figure~\ref{fig:panel_nex_fq}.
Indeed, that is broadly the case; in both samples the power spectra exhibit a slow increase with decreasing angular scale until the power spectra turnover at small scales.  Cosmic variance appears to be significantly reduced for $\mathrm{\breve{f}{q}}$ even in the spectroscopic case.   Good agreement is obtained between the observed and simulated correlations, despite the fact that quenching is defined in terms of $\mathrm{(u-r)_{obs}}$ colour in the photometric sample and sSFR in the spectroscopic sample for the observations.  (One might have worried that dusty red, but star-forming galaxies might have contaminated the colour-based quenched fraction at some level, but that does not appear to be the case.)  BAHAMAS, which uses sSFR for both the shallow ($z < 0.06$) and deeper ($z < 0.15$) samples to determine quenched fraction, reproduces the observed clustering of quenched fraction remarkably well.

EAGLE exhibits an earlier turn off at $\ell~\sim 600$ in panel ($d$) of Fig.~\ref{fig:panel_nex_fq}, compared to $\ell~\sim 1000$ for BAHAMAS and SDSS.   Note that no such feature is visible in the EAGLE galaxy power spectrum at this scale (panel $b$), which rules out a difference in smoothing origin.  We, therefore, conclude that this is a genuine issue.  More generally, it is interesting that the clustering of quenched fraction (panels $c$ and $d$) drops off at small scales faster than does the clustering of galaxies in general (panels $a$ and $b$).   As just mentioned, this cannot be a result of differences in smoothing, as the quenched fraction and galaxy overdensity have been smoothed in exactly the same way.  Na\"ively, one might have expected the opposite trend (i.e., that quenching becomes more prevalent on small scales).  However, it should be kept in mind that the contribution of different types of systems can vary depending on the particular auto- and cross-spectrum being examined, as well as the scales under consideration.  For example, the fact that the degree of cosmic variance in the simulations is relatively large for the galaxy overdensity power spectrum implies that it is dominated by relatively rare systems (e.g., clusters).  The quenched fraction power spectrum, on the other hand, shows little variation from cone to cone (even for the spectroscopic selection using EAGLE), which strongly suggests that this correlation is dominated by relatively common objects (e.g., central galaxies near the lower mass limit of the sample).  We discuss this further in Section \ref{section:discuss}.   

In a future study, we will examine these issues in greater detail, by deconstructing the simulation power spectrum into contributions from, e.g., galaxies of different stellar mass, halo mass, and redshift, centrals vs. satellites, host halo mass, and so on.

\subsection{Quenched fraction--galaxy overdensity cross-spectrum}

As mentioned above, galaxy density is known to correlate with environmental quenching. Can we measure this correlation with our method?  Panels ($e$) and ($f$) of Figure~\ref{fig:panel_nex_fq} show the quenched fraction--overdensity ($\mathrm{\breve{f}_q - \breve{N}_{tot}}$) cross-spectrum. These strong detections confirm that, indeed, quenched fraction and galaxy overdensity are spatially correlated. The cross-spectrum has a very similar shape to the auto-spectra of its constituents: gradually rising at low multipoles, plateauing and then rapidly dropping to zero at small scales.  The scales at which the cross-power spectra turn over are intermediate to those seen the galaxy overdensity and quenched fraction power spectra.  Interestingly, the degree of cosmic variance (cone-to-cone scatter) in the quenched fraction--overdensity cross-spectrum is significantly larger than for the quenched fraction power spectrum.  This suggests that this correlation is picking out a population that is relatively rare (e.g., associated with massive systems).  Indeed, we will show in Section \ref{section:discuss} that this cross-spectrum is particularly sensitive to the quenching of satellite galaxies, whereas the quenched fraction power spectrum (auto-correlation) is significantly less so.

Relatively good agreement is obtained between observations and both of the simulations, although the overall shape and amplitude are by no means perfectly reproduced.  

\subsection{tSZ effect--galaxy overdensity cross-spectrum}

\begin{figure*}
\includegraphics[width=0.97\linewidth]{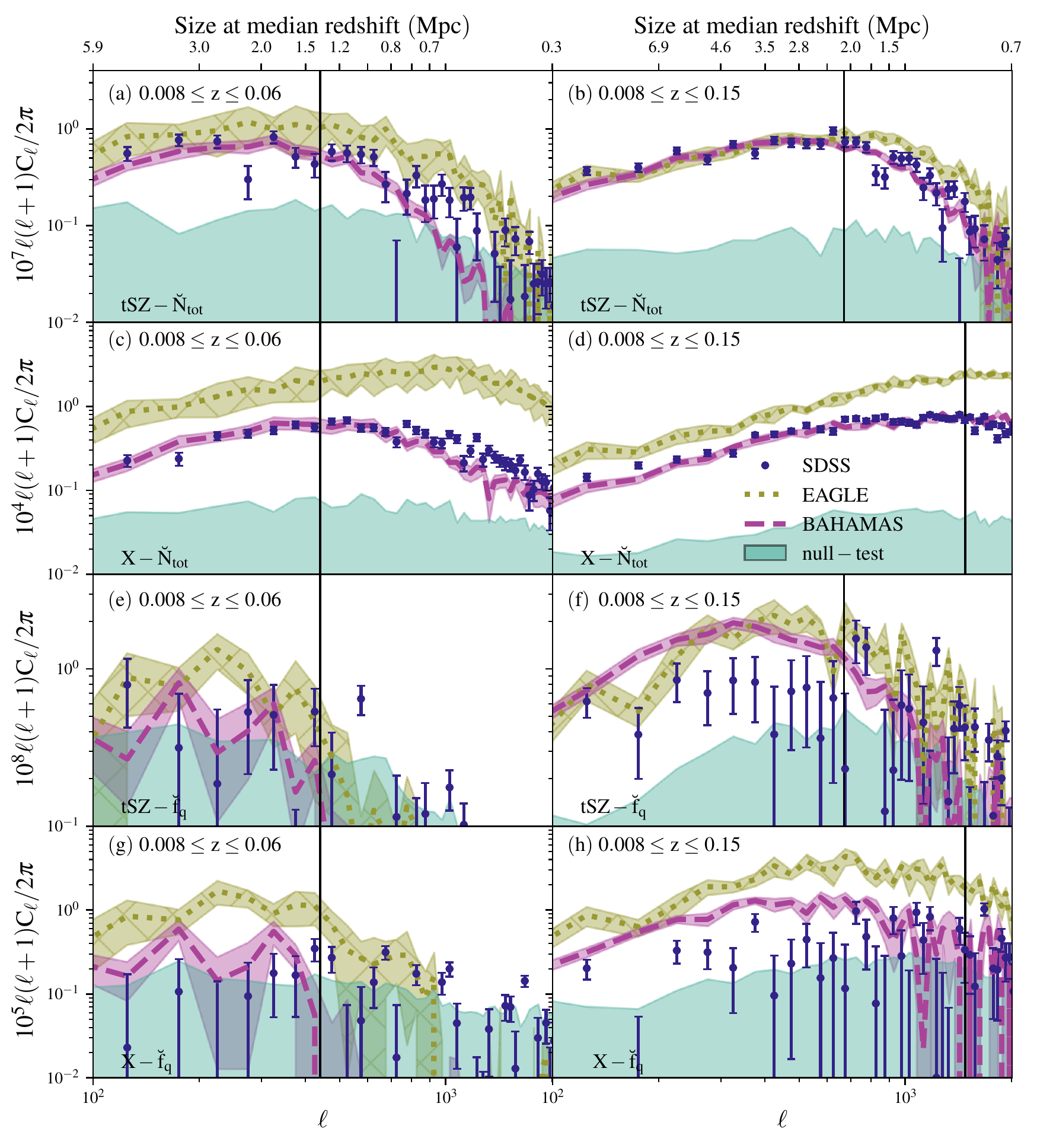}
\caption{Cross-power spectra between galaxy properties and hot gas measures, for the spectroscopic sample (\emph{left}) and photometric sample (\emph{right}).  Navy data points with error bars correspond to observed cross-spectra, dashed and dotted curves correspond to the predictions BAHAMAS and EAGLE, respectively.  The shaded regions around the curves show $1\sigma$ confidence interval on the mean, while the turquoise regions at the bottom of each panel shows the $1\sigma$ level scatter around zero from 100 null-tests.  Vertical, black lines indicate the scale at which SPH-smoothed power spectra deviate from their discrete counterparts by 50\% due to either SPH smoothing or $Planck$ beam effects (see Appendix \ref{appendix:sph_smoothing}).  The top x-axis shows the approximate physical scale assuming the median redshift of each sample. The top row (panels~($a$) and ($b$)) shows the tSZ effect--galaxy overdensity ($\mathrm{tSZ} - \mathrm{\breve{N}_{tot}}$) cross-power spectra.  The second row (panels~($c$) and ($d$)) shows the X-ray flux--galaxy overdensity ($\mathrm{X} - \mathrm{\breve{N}_{tot}}$) cross-spectra. The third row (panels~($e$) and ($f$)) shows the tSZ effect--quenched fraction excess ($\mathrm{tSZ} - \mathrm{\breve{f}_{q}}$) cross-spectra.  The bottom row shows the X-ray flux--quenched fraction excess ($\mathrm{X} - \mathrm{\breve{f}_{q}}$) cross-spectra.  Strong detections are made for the hot gas--galaxy overdensity cross-spectra for both the spectroscopic and photometric selection.  The hot gas--quenched fraction excess cross-spectra, on the other hand, are only well measured for the deeper photometric selection.  Both EAGLE and BAHAMAS reproduce the tSZ--overdensity cross-spectra, but EAGLE predicts a larger than observed amplitude for the X-ray flux--overdensity cross-spectra (see text).  Both simulations predict hot gas--quenched fraction cross-spectra that are higher in amplitude than observed, likely as a result of overly efficient quenching of satellite galaxies (see Section \ref{section:discuss}).}
\label{fig:panel_tsz_x}
\end{figure*}

Having established that a measurable signal can be obtained from correlations in galaxy properties alone, with overdensity being a commonly-used proxy for environment, we now turn our attention to direct environmental measures, specifically, the hot gas properties. Panels ($a$) and ($b$) of Figure~\ref{fig:panel_tsz_x} show the cross-spectrum between the tSZ effect $y$ parameter and galaxy overdensity.

Since the $Planck$ tSZ effect maps are convolved with a 10 arcmin beam, we do the same for the simulated tSZ effect maps.  The scale of this beam, $\mathrm{\theta_{FWHM} = 10'}$, corresponds to the angular scale of $\ell = 1080$, however, the effects become apparent on significantly larger scales. In the same way as for SPH smoothing effects, we indicate the scale at which beam-convolved power spectra deviate from beam-free by 50\%. This occurs at $\ell = 670$. The small-scale decline in power is now dominated by $Planck$ beam in the `photometric' case. 

An examination of the spectroscopic case in panel ($a$) reveals that even with a shallow galaxy sample a strong signal can be measured. Good agreement is achieved between the observations and BAHAMAS, whereas EAGLE predicts a slightly stronger correlation than is observed. Given the cosmic variance present in the spectroscopic case of $\mathrm{\breve{N}_{tot} - \breve{N}_{tot}}$ in panel ($a$) of Fig.~\ref{fig:panel_nex_fq}, it is reasonable to expect the same here.  Indeed, the scatter between individual light cones is sufficiently large to account for the discrepancies between the observations and simulations. The power spectra are even biased in the same way: EAGLE over-estimates the power on all scales, especially at high $\ell$; BAHAMAS agrees with observations for all but the smallest scales where it under-predicts the signal slightly.

Just as in the galaxy overdensity power spectrum ($\mathrm{\breve{N}_{tot} - \breve{N}_{tot}}$) case, cosmic variance is greatly reduced in the photometric sample (panel $b$), where the measured cross-power spectra agree rather well between both simulations and observations.

\subsection{X-ray--galaxy overdensity cross-spectrum}

Panels ($c$) and ($d$) of Fig.~\ref{fig:panel_tsz_x} show the analogous cross-correlation for X-ray photon flux (as opposed to tSZ effect).  While both the tSZ effect and X-ray flux are associated with the same hot gas, their connection to quenching could be quite different.  However, there is no evidence for a qualitatively different correlation with galaxy overdensity.  With the exception of normalisation, the cross-spectrum profiles look very similar for X-ray and tSZ effect. 

EAGLE is once again significantly higher than the observations and BAHAMAS. EAGLE's high amplitude, which is present in the photometric selection as well (and is therefore not due to cosmic variance), is expected in this case.  It is already established that the AGN feedback present in EAGLE REF simulation is not sufficiently strong to remove an appropriate amount of gas from galaxy groups (see \citealt{Schaye_etal_2015}). Haloes, therefore, contain too much hot gas, which leads to excessive X-ray luminosities, as reported recently by \citet{Davies_etal_2019}.  While the ratio of EAGLE to observed X-ray luminosities \citep{Anderson_etal_2016} at fixed halo mass is $\sim 4$, it requires more complex modelling to introduce a correction factor into the cross-power spectrum with overdensity.  As it is not the subject of this paper, we simply report the measured signal as it is measured.

Despite the noisy nature of X-ray observations, strong detections are made for both samples.  With the two simulations in hand, one of which agrees with observations while the other does not, it should be possible to decompose the signals and identify the physical cause (e.g., differences in feedback) that  lead to these differences.   This, in turn, should shed light on how exactly gas density/pressure are connected to the quenching of galaxies. This needs to be investigated in the future.

Note that the simulated X-ray maps contain emission from hot gas only, whereas the RASS X-ray map also contains point sources (e.g., AGN, X-ray binaries, stars, etc.) which we have not masked out.   While X-ray AGN are more prevalent (by abundance) at high redshifts (e.g., \citealt{Miyaji2001,Hasinger2005}), they do exist locally as well and have been shown to spatially trace the normal galaxy population (e.g., \citealt{Krumpe2012}).  We might therefore expect them to contribute to the observed X-ray cross-spectra and to potentially bias the comparison with the simulations, which do not model this effect.  In Appendix~\ref{appendix:AGN} we have explicitly checked the level of bias present in the recovered cross-spectra, by masking out AGN in the RASS point source catalogue and recomputing the observed cross-spectra.  We find the level of bias present to be small (generally resulting in less than a 1-sigma change to the measurements at individual multipoles), such that none of the main conclusions of the present study are affected by neglecting the contribution of X-ray AGN.

\subsection{tSZ effect--quenched fraction cross-spectrum}

We present the cross-correlation power spectra between quenched fraction and tSZ effect $y$ parameter in panels ($e$) and ($f$) of Fig.~\ref{fig:panel_tsz_x}.  Importantly, this correlation is independent of galaxy overdensity or whether a group/cluster is present.  It is simply asking whether the quenching of galaxies knows about the local hot gas conditions.  It is immediately evident that the signal-to-noise ratio of this cross-correlation is significantly lower than all the previous cases.  

Visual inspection of panel ($e$) of Fig.~\ref{fig:panel_tsz_x} reveals that only the large scale contributions are (marginally) detected, with practically no signal present above $\ell\sim500$. This is in line with all detected quenched fraction signals in this redshift bin; at that scale $\mathrm{\breve{f}_{q} - \breve{f}_{q}}$ power spectrum in panel ($c$) of  Fig.~\ref{fig:panel_nex_fq} is very much on the decline and $\mathrm{\breve{f}_{q} - \breve{N}_{tot}}$ in panel ($e$) of Fig.~\ref{fig:panel_nex_fq} is similarly close to zero.  The simulations more or less support this trend, although with a large degree of scatter.

Using the deeper photometric sample (panel $f$ of Fig.~\ref{fig:panel_tsz_x}), the tSZ effect--quenched fraction cross-spectrum is detected on most scales.  The simulations produce similar correlations to each other over the entire range of scales, as in panel ($b$) of Fig.~\ref{fig:panel_tsz_x}, however they both predict amplitudes that are too high relative to the observed tSZ effect--quenched fraction cross-spectrum.  We discuss possible reasons for this difference below, in Section \ref{section:discuss}.

\subsection{X-ray--quenched fraction cross-spectrum}

Finally, we examine the X-ray--quenched fraction (X-ray$-\mathrm{\breve{f}_{q}}$) cross-power spectra in panels ($g$) and ($h$) of Fig.~\ref{fig:panel_tsz_x}.   The spectroscopic galaxy sample is not sufficiently deep to measure this signal for any analysis. While these detections are weak, a general behaviour of the correlation can still be seen, especially so for the deeper (photometric) of the two samples.

The trend of simulations overestimating the signal seen in panel ($f$) of Fig.~\ref{fig:panel_tsz_x} is also present here. EAGLE shows the same excess in signal relative to BAHAMAS as is seen in X-ray$-\mathrm{\breve{N}_{tot}}$ cross-correlation, which can be attributed to excessive X-ray luminosities in the former.  Both simulations, however, predict cross-spectra that are in excess of what is observed. 

\section{Discussion: isolating external from internal quenching}
\label{section:discuss}

As the simulations (particularly BAHAMAS) yield a reasonable match to the observed correlations, we can use them to gain some further insight into the physical drivers of the observed correlations presented above.  We leave a detailed deconstruction of the auto- and cross-spectra for future work (Kukstas et al., in prep), commenting here only on the respective roles of internal and external quenching.  In particular, thus far we have not made any distinction between central and satellite galaxies when dealing with sample selection, map making, or cross-correlation, in either the observations or simulations.  This is partly due to the fact that this is a non-trivial task for observations, particularly those based on photometric redshifts.  Here we note that upcoming large optical surveys (LSST, Euclid) will be photometric only.  However, we can easily separate simulated galaxies into centrals and satellites (as well as by a wealth of other available information) and see what this does to the predicted correlations.

\begin{figure}
	\includegraphics[width=\columnwidth]{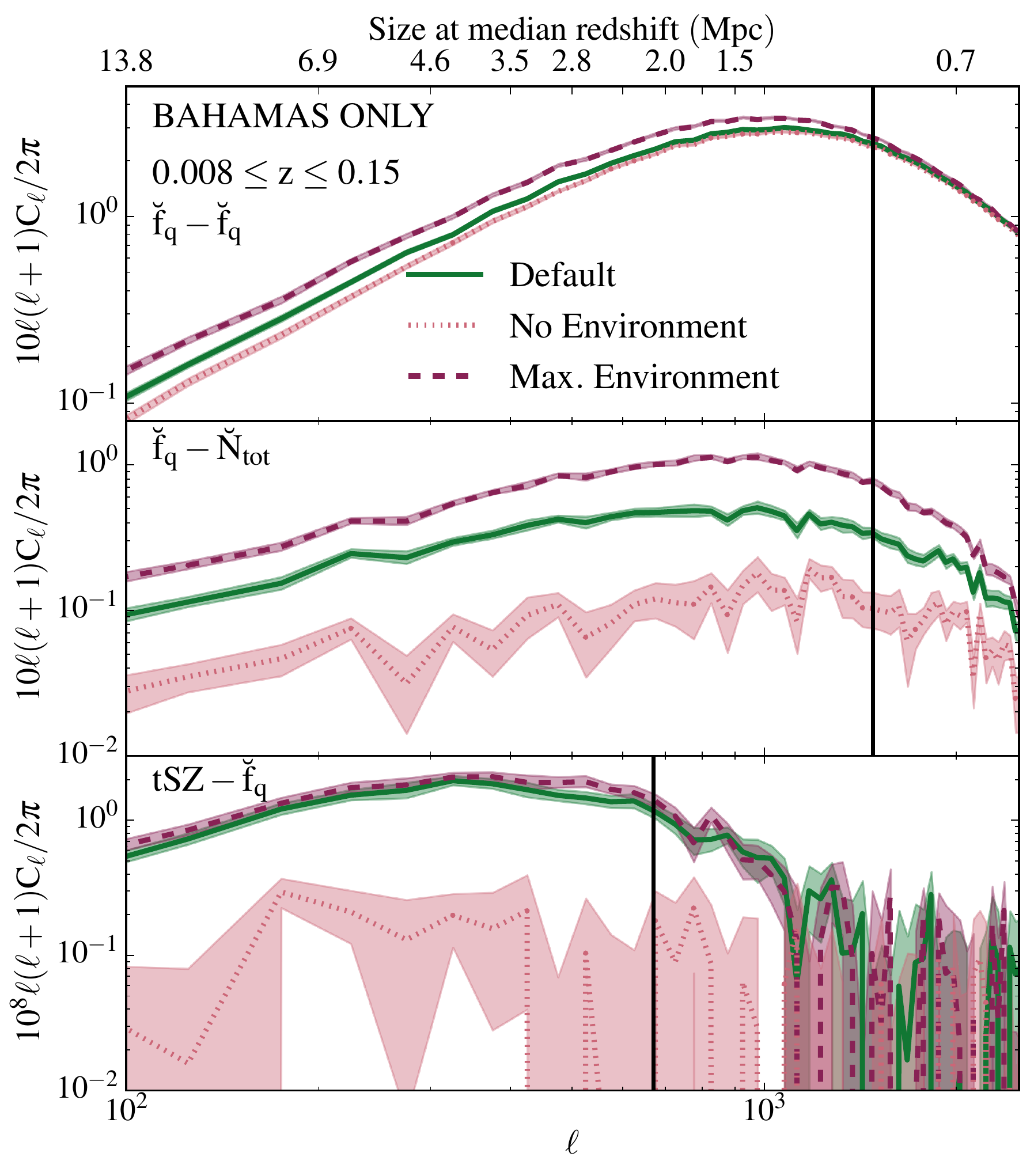}
    \caption{The impact of changing the satellite quenched fraction to either fully quenched (`max environment') or so that it statistically matches that of centrals (`no environment') for BAHAMAS galaxies in the photometric sample.  The solid curves represent the default (unmodified) power spectra shown in the previous plots.  The dashed curves correspond to the case where all satellites are quenched and the dotted curves represent the case where satellite quenched fraction is adjusted to match centrals.
    \emph{Top}: quenched fraction power spectrum.  \emph{Middle}: quenched fraction--overdensity cross-spectrum.  \emph{Bottom}: tSZ effect--quenched fraction cross-spectrum.  The quenched fraction--overdensity and (particularly) the tSZ effect--quenched fraction cross-spectrum are particularly sensitive to the nature of satellite quenching, whereas the quenched fraction power spectrum (auto) is only mildly sensitive (being driven mainly by internal/mass quenching).}
    \label{fig:change_sat_fq}
\end{figure}

In order to test the sensitivity of the measured signals to internal and external quenching, we explore the extremes of satellite quenched fraction.  In particular, we artificially change the specific star formation rates of satellites in the BAHAMAS simulation such that they are either (1) all quenched; or (2) match the $\mathrm{f_q - log_{10}(M_*)}$ relation of centrals.   (To achieve the latter, we randomly sample the sSFR distribution of central galaxies in a $\mathrm{log_{10}(M_*)}$ bin and assign sSFRs to a given satellite in the same bin.)  This results in two samples where: (1) the environmental effects are maximally efficient (all satellites are quenched); or (2) they are non-existent (satellites are statistically the same as centrals).  It is important to note that the population of centrals is unchanged in this process and the total galaxy density remains unaltered in the maps.  To make the signals directly comparable to those previously measured, we have also used the same mean value $\mathrm{f_q}$ in computing the quenched fraction excesses.  Thus, everything is measured relative to the default case presented in the previous plots. Fig.~\ref{fig:change_sat_fq} shows a selection of cross-correlations taken from BAHAMAS where satellite quenched fraction has been modified as described above.

The top panel contains the quenched fraction power spectrum ($\mathrm{\breve{f}_q}$--$\mathrm{\breve{f}_q}$).  The solid, green curve is the unmodified power spectrum from panel ($b$) of Fig.~\ref{fig:panel_nex_fq}, the maroon, dashed curve is for central-matched-$\mathrm{f_q}$ (i.e., no external quenching), and coral, dotted curve represents the maximum environmental quenching case.  While the quenched fraction power spectrum is sensitive in detail to external/environmental quenching (particularly on large scales), it is clearly mostly driven by internal quenching, as varying the external quenching in extreme ways only produces a relatively mild effect on the predicted power spectrum.

This behaviour is in strong contrast to the quenched fraction--galaxy overdensity cross-spectrum (middle panel), where the predicted signal is increased by a factor of $\sim2$ (relative to the default case) when all satellites are quenched, and decreased by a factor of $\sim4$ when environmental effects are completely absent.  

Finally, in the bottom panel we show the predicted tSZ effect--quenched fraction cross-spectra for the different environmental quenching cases.  This correlation is incredibly sensitive to the nature of external/satellite quenching: when satellite quenching is turned off the cross-spectrum is reduced by approximately an order of magnitude.  The fact that the maximum external quenching is so similar (though not identical) to the default case, suggests that satellite quenching is particularly strong in the simulations.  This is true for the EAGLE simulations as well (not shown).  Since both BAHAMAS and EAGLE predict tSZ effect--quenched fraction cross-spectra that are in excess of the observed cross-spectrum from SDSS and $Planck$, this suggests that satellite quenching in the simulations, particularly in relation to the local hot gas properties, is too efficient.  Interestingly, this conclusion seems consistent with the recent findings of \citet{Bahe_etal_2017_hydrangea}, who used the Hydrangea zoomed simulations of galaxy clusters (which were run with the EAGLE code) to examine the efficiency of satellite quenching with respect to the optical group catalogue-based findings of \citet{Wetzel_2012} (see figure 6 of \citealt{Bahe_etal_2017_hydrangea}).  (See also \citealt{Lotz_etal_2019} for similar conclusions based on the Magneticum Pathfinder simulations.)
Ascertaining why the simulations are too efficient at quenching satellites should be a high priority.

\section{Summary and Conclusions}
\label{section:conclusions}

The primary goal of this paper is to establish a new methodology, based on spatial cross-correlations, for testing physical models of the environmental-dependence of galaxy evolution.  In this regard, our study is very much a proof of concept.  We have established that these auto- and cross-spectra can be detected and measured even in current data (and will therefore yield \emph{booming} signals in future surveys, e.g., LSST, Euclid, Simons Observatory, CMB-S4, eROSITA) and that state-of-the-art simulations yield a reasonable match to most, but not all, of the observed correlations.  We also demonstrated that different power and cross-power spectra are sensitive to internal (e.g., AGN and stellar feedback) and external (e.g., ram pressure stripping, harassment, strangulation, etc.) quenching in different ways, allowing one in principle to constrain models for both simultaneously.

Below we summarise the main points:

\begin{itemize}
    \item We constructed two volume-limited and stellar mass-complete samples from the SDSS.  One is based on the DR7 spectroscopic sample ($z < 0.06$) and the other on the DR12 photometric sample ($z < 0.15$) - see Fig.~\ref{fig:stellar_mass_fn}.  Specific star formation rates (sSFRs) and colours were used to assign a star-forming/quenched status for the two samples, respectively (see Figs.~\ref{fig:colour_ssfr_correlation} and \ref{fig:fq_logMstar}).
    \item The SDSS samples were projected onto \texttt{HEALPix} images, to create maps of galaxy overdensity and quenched fraction overdensity.  These are then used to compute angular power spectra and cross-spectra together with maps of the thermal Sunyaev-Zel'dovich effect from $Planck$ and diffuse X-ray emission from the ROSAT All-Sky Survey.  See maps in Fig.~\ref{fig:maps}.   
    \item We used the publicly-available software package \verb'NaMaster' to compute the auto- and cross-power spectra from the \texttt{HEALPix} maps.
    \item Strong detections are reported for the auto- and cross-spectra involving galaxy properties only, including the galaxy overdensity power spectrum, the quenched fraction power spectrum, and the galaxy overdensity--quenched fraction cross-spectrum (see Fig.~\ref{fig:panel_nex_fq}).  Of these correlations, the galaxy overdensity--quenched fraction cross-spectrum is particularly sensitive to satellite quenching, whereas the quenched fraction power spectrum is considerably less so (see Fig.~\ref{fig:change_sat_fq}).
    \item Using synthetic observations of the EAGLE and BAHAMAS simulations, we show that, overall, the simulations reproduce the auto- and cross-spectra involving galaxy properties alone reasonably well, although they tend to predict a larger than observed amplitude for the overdensity--quenched fraction cross-spectrum.  This suggests that satellite quenching may be too efficient in the simulations.
    \item Strong observational detections are also reported for the cross-spectra involving galaxy overdensity and either tSZ effect or X-ray surface brightness (see top panels of  Fig.~\ref{fig:panel_tsz_x}).  The BAHAMAS simulations reproduce these cross-spectra well, whereas the EAGLE simulations predict a larger than observed amplitude for the X-ray--overdensity cross-spectrum, likely due to inefficient feedback on the scale of groups.
    \item We also report, for the first time, detections of the quenched fraction--tSZ effect and quenched fraction--X-ray flux cross-spectra (see bottom panels of Fig.~\ref{fig:panel_tsz_x}).  No information about galaxy overdensity or the presence of known galaxy groups/clusters is used here.
    \item Both BAHAMAS and (particularly) EAGLE predict larger than observed amplitudes for the quenched fraction--tSZ/X-ray cross-spectra.  As these cross-spectra are remarkably sensitive to the nature of satellite quenching (see Fig.~\ref{fig:change_sat_fq}), these results again suggest that the quenching of satellites in the simulations, particularly in relation to local hot gas properties, is too efficient in the simulations.
\end{itemize}

In a future study, we plan to examine the theoretical predictions in more detail, by deconstructing the power and cross-power spectra into contributions from, e.g., galaxy stellar mass, halo mass, redshift, central/satellite designation, host halo mass for satellites, and so on.  This should yield further insight into the successes and failures of the simulations in reproducing the observed correlations reported here.

Finally, we point out that our methodology is not limited to linking galaxy quenched fractions to hot gas properties.  One can easily substitute out quenched fraction for any galaxy-based property (e.g., a morphology-based quantity such as disk-to-total ratio, S\'ersic index, or concentration) and/or substitute out hot gas properties for some other direct measure of environment, such as weak lensing potential.  The advent of wide field galaxy and large-scale structure surveys now present us with a multitude of ways to directly link galaxies with their environments and this will only improve in the coming years with new surveys coming online.  The increased statistics should also allow one to explore what we term `environmental tomography' (in analogy to cosmic shear tomography), whereby the auto- and cross-spectra can be evaluated in redshift bins, to probe the redshift evolution of the correlations and also achieve a closer-to-3D view of the local physical environment.  Such data, when compared carefully to simulations, holds the promise of developing a detailed physical picture for the environmental evolution of galaxies.

\section*{Acknowledgements}

The authors thank the EAGLE team for making their simulations publicly available.  They also thank Joop Schaye for his contributions to the BAHAMAS simulations. EK thanks Rob Crain, Chris Collins, and Lee Kelvin for helpful discussions, and Yannick Bah\'e for insightful comments.

This project has received funding from the European Research Council (ERC) under the European Union's Horizon 2020 research and innovation programme (grant agreement No 769130).  

This work used the DiRAC@Durham facility managed by the Institute for
Computational Cosmology on behalf of the STFC DiRAC HPC Facility
(\url{www.dirac.ac.uk}). The equipment was funded by BEIS capital funding
via STFC capital grants ST/K00042X/1, ST/P002293/1, ST/R002371/1 and
ST/S002502/1, Durham University and STFC operations grant
ST/R000832/1. DiRAC is part of the National e-Infrastructure.

We acknowledge the use of data provided by the Centre d'Analyse de Donn\'ees Etendues (CADE), a service of IRAP-UPS/CNRS (http://cade.irap.omp.eu, Paradis et al., 2012, A\&A, 543, 103).

This work is partly based on observations obtained with $Planck$ (http://www.esa.int/Planck), an ESA science mission with instruments and contributions directly funded by ESA Member States, NASA, and Canada.

GAMA is a joint European-Australasian project based around a spectroscopic campaign using the Anglo-Australian Telescope. The GAMA input catalogue is based on data taken from the Sloan Digital Sky Survey and the UKIRT Infrared Deep Sky Survey. Complementary imaging of the GAMA regions is being obtained by a number of independent survey programmes including GALEX MIS, VST KiDS, VISTA VIKING, WISE, Herschel-ATLAS, GMRT and ASKAP providing UV to radio coverage. GAMA is funded by the STFC (UK), the ARC (Australia), the AAO, and the participating institutions. The GAMA website is http://www.gama-survey.org/ .

\emph{Software:} This work was made possible due to a large number of open-source software packages, including PYTHON \citep{Python_2011}, IPYTHON \citep{IPython_2007}, ASTROPY \citep{astropy_2013}, MATPLOTLIB \citep{Hunter_2007}, NUMPY \citep{vdWalt_2011}, SCIPY \citep{Jones_2001}, HEALPY \citep{Zonca_2019}.




\bibliographystyle{mnras}
\bibliography{refs} 



\appendix

\section{SPH smoothing and shot noise}
\label{appendix:sph_smoothing}

In order to map a point object, such as a galaxy from a catalogue or a gas particle from a simulation, onto an extended grid, we make use of a kernel interpolation technique commonly used in smoothed particle hydrodynamic simulations. The kernel, $\mathrm{W(r,h)}$, is defined as a spline with a smoothing scale $h$:
\begin{equation}
    W(r,h)=\frac{8}{\pi h^3}
    \begin{cases}
    ~1-6~(\frac{r}{h})^2+6~(\frac{r}{h})^3 ,& 0 \leq \frac{r}{h} \leq \frac{1}{2},\\
    ~2~(1-\frac{r}{h})^3,                  & \frac{1}{2} < \frac{r}{h} \leq 1,\\
    ~0,                                   & \frac{r}{h} > 1.
\end{cases}
\end{equation}
In the case of simulations, $h$ is determined by the size of a gas particle's 3D smoothing length, whereas when making galaxy maps $h$ is determined by:

\begin{equation}
    h = \sqrt{\frac{N_{sph}}{\pi n}},
\end{equation}
where $\mathrm{N_{sph}}$ is the number of nearest neighbours to be smoothed over and $\mathrm{n}$ is the mean surface density of galaxies for the given map.

\begin{figure}
\begin{subfigure}[t]{.475\textwidth}
	\includegraphics[width=\linewidth]{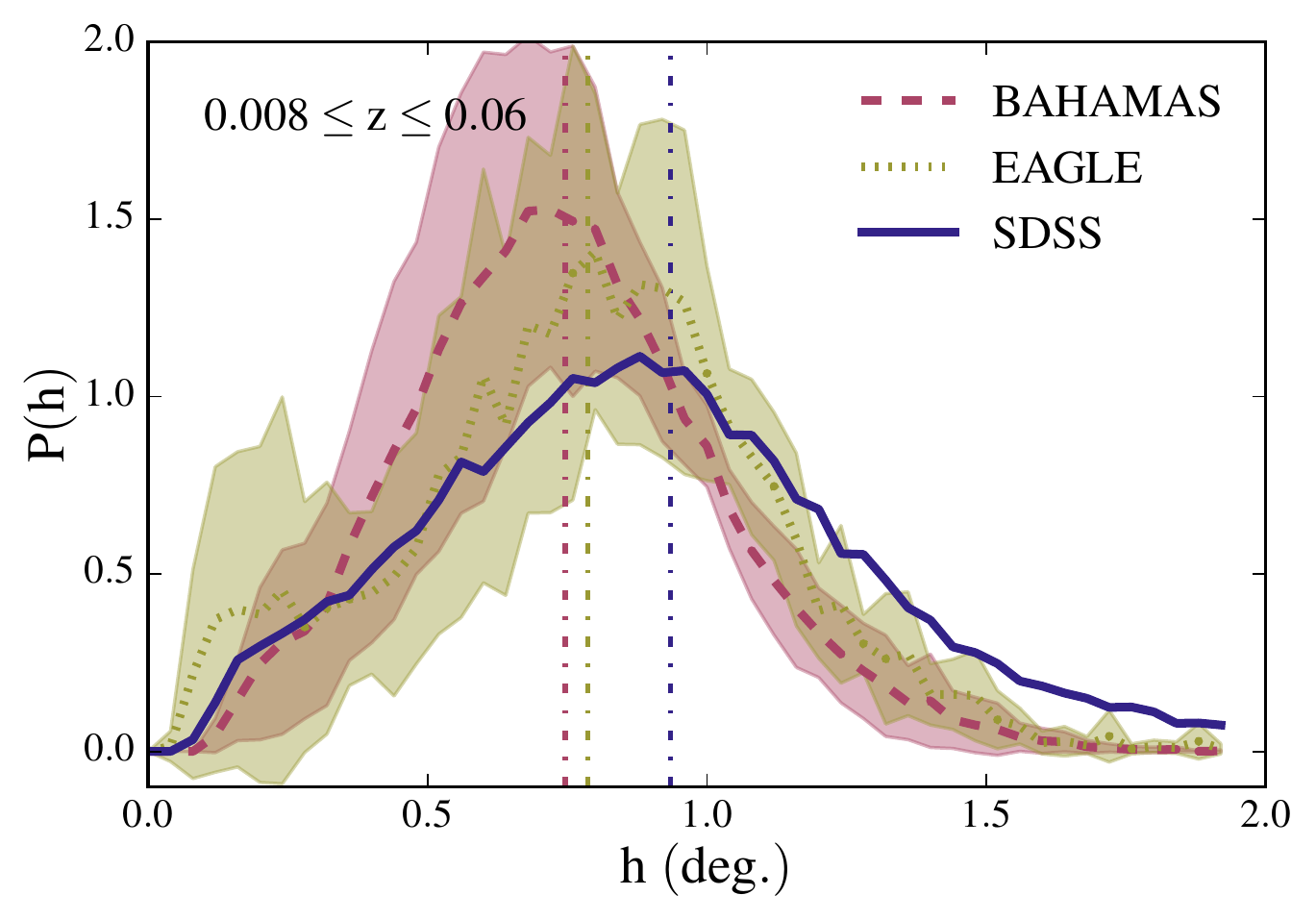}
    \caption{Distribution of smoothing parameter ($h$) for the EAGLE, BAHAMAS, and SDSS photometric low-redshift samples. Vertical, dash-dotted lines indicate the mean smoothing scale for each case. While the distributions are similar in their shapes and mean smoothing scale, differences will translate into smalle-scale (high $\mathrm{\ell}$) power-spectrum deviations.}
    \label{fig:sph_h_lowz}
\end{subfigure}\\

\begin{subfigure}[t]{.475\textwidth}
	\includegraphics[width=\columnwidth]{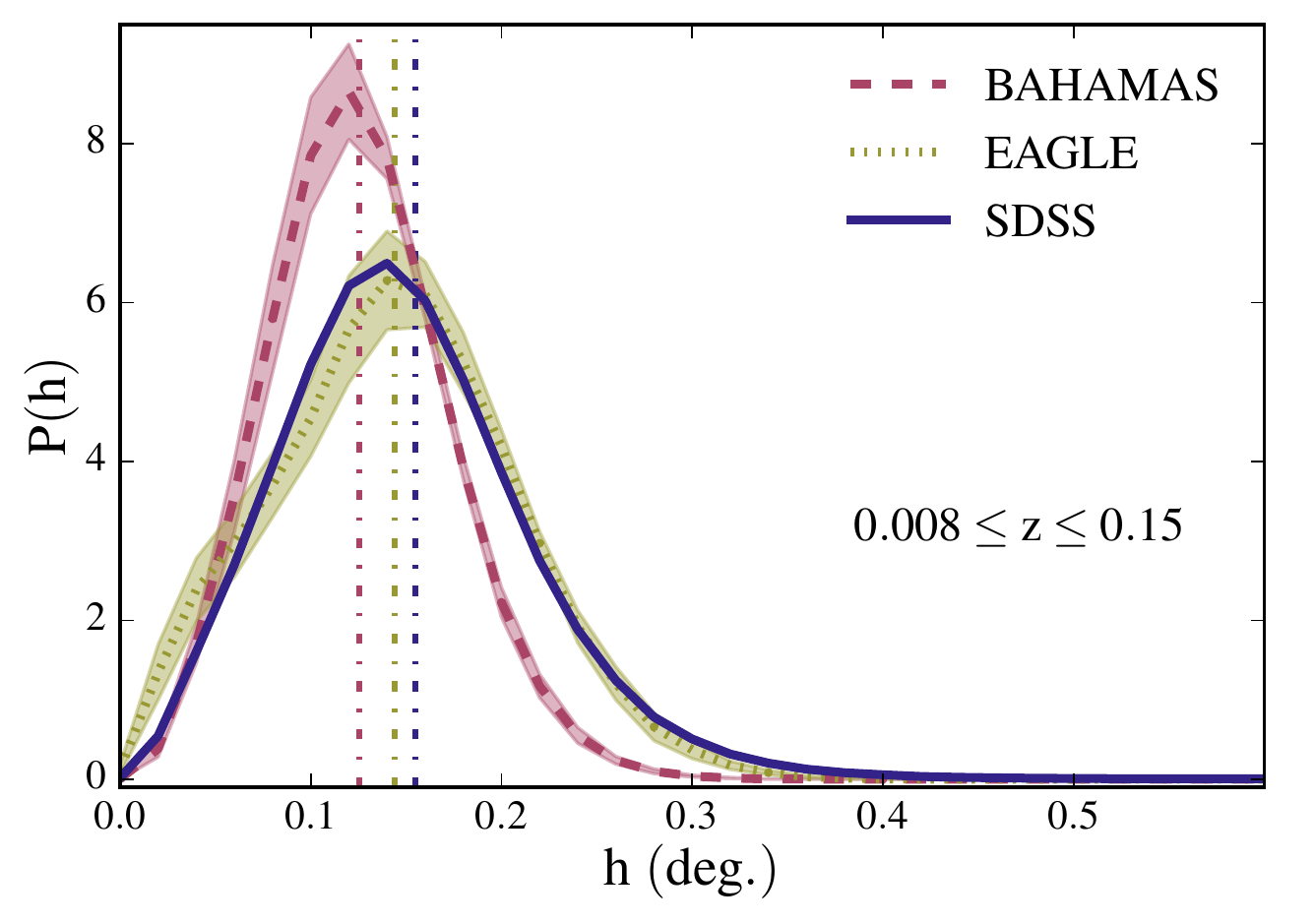}
    \caption{The same as Fig.~\ref{fig:sph_h_lowz} but for the high-redshift sample. Note the shift to smaller smoothing scales. }
    \label{fig:sph_h_highz}
\end{subfigure}
\caption{}
\end{figure}

This spline can be approximated with a Gaussian function, with one notable exception that the spline has a well-defined extent. The distributions of $h$ for each simulation/map can be seen in Figures~\ref{fig:sph_h_lowz} and \ref{fig:sph_h_highz}. All samples show similar distributions in $h$, with `characteristic scales' similar between corresponding samples in simulations and observations. The difference in number of galaxies is reflected in the size of each kernel, with spectroscopic sample demanding a kernel as large as 2 degrees and photometric sample requiring one no larger than 0.4 deg.

\begin{figure}
        \includegraphics[width=\linewidth]{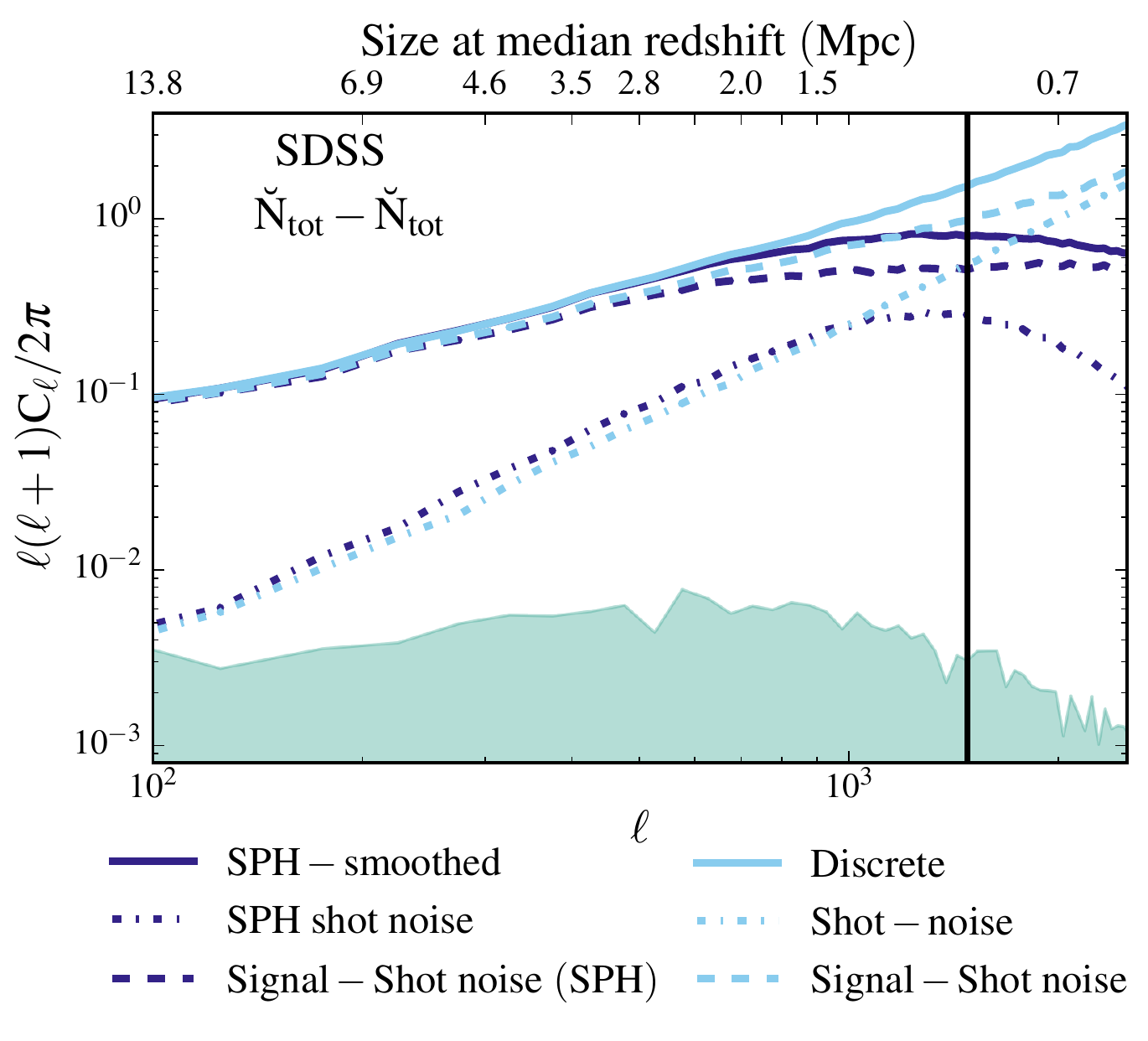}
    \caption{A demonstration of SPH smoothing and shot noise on the power spectra for the photometric galaxy sample. Solid lines represent $\mathrm{\breve{N}_{tot}}$-auto power spectra for discrete (cyan) and SPH-smoothed (navy) cases. Solid, black, vertical line indicates the $\mathrm{\ell}$-scale where they begin to differ by more than 50\%. Dash-dotted lines show the level of shot noise present in our galaxy maps in discrete (cyan) and SPH-smoothed (navy) forms. Dashed lines show the shot noise-subtracted power spectra in both cases.}
    \label{fig:shot_noise}
\end{figure}

The overall effect of smoothing on a power spectrum can be seen in Figure~\ref{fig:shot_noise} as solid lines, where two versions of $\mathrm{\breve{N}_{tot} - \breve{N}_{tot}}$ (total galaxy overdensity) power spectrum are presented; before smoothing (cyan) and after (navy).  The discrete (i.e., unsmoothed) map used to compute this power spectrum was constructed by assigning a value of zero to all pixels with no galaxies within the main SDSS footprint, rather than using SPH smoothing to interpolate.   (Note that area outside the footprint is masked as shown in Fig.~\ref{fig:maps} [panel $d$].)  The unsmoothed power spectrum rises until it reaches the pixel scale and drops abruptly to zero, which occurs at angular scales significantly smaller than smoothing scales discussed in this paper.  

In order to indicate a characteristic scale of SPH smoothing effects, we plot a vertical, black solid line at the scale which marks a $50\%$ deviation of SPH-smoothed power spectrum from the discrete case.  For SPH smoothing, these scales are: $\mathrm{\ell = 670}$ and $\mathrm{\ell = 1480}$ for the `spectroscopic' and `photometric' samples, respectively.  This is relative to $\mathrm{\ell = 440}$ and $\mathrm{\ell = 3700}$ for $Planck$ and ROSAT beam effects, respectively.  Therefore, the power spectra are limited by SPH smoothing scale in all cases except tSZ cross-correlations with the 'photometric' sample, where $Planck$ beam effects dominate. The vertical line indicates this accordingly. The effects of SPH smoothing are the same for both simulations, within cosmic variance uncertainty.

Figure~\ref{fig:shot_noise} also shows the effects of shot noise on the galaxy overdensity power spectrum.  The dash-dotted lines show the level of shot-noise present in the maps.  These were computed following \citet{Feldman_etal_1994}, by randomising galaxy position coordinates in order to remove any structure present in the maps. This yields a power spectrum of constant $\mathrm{C_{\ell}}$ over the scales presented here.  The shot noise power spectrum responds to SPH smoothing in the same way as the signal (cyan and navy dash-dotted lines). We also checked that the effect is the same in the simulations but do not show this explicitly. All $\mathrm{\breve{N}_{tot}}$ power spectra presented in the main text are shot noise subtracted.

How to rigorously estimate the shot noise contribution to the $\mathrm{\breve{f}_{q} - \breve{f}_{q}}$ correlation (quenched fraction power spectrum) is less obvious, however.  Since the mean quenched fraction is not conserved when making randomised maps (there is no information about the \emph{number} of galaxies used to compute quenched fraction), it is unclear how to calculate a normalisation for the shot noise contribution.  We therefore leave the quenched fraction power spectrum affected by shot noise.  However, we do not expect shot noise to be a dominant component for this power spectrum for two reasons.  First, shot noise is subdominant in both the total galaxy density and quenched galaxy density power spectra and (ignoring the differences of slightly different smoothing kernels)) if $\mathrm{\breve{f}_{q}}$ is just a division of the two, then its shot noise properties would also be sub-dominant. Secondly, as demonstrated by computing the null-tests, the relative orientation of the two maps is very important. The measured signal vanishes if the maps are mis-aligned, this would not be the case if they were shot noise-dominated.

Lastly, we highlight that shot noise does not affect any cross-spectra, as noise properties are different and certainly not spatially correlated between two different maps.

\section{Contamination from AGN}
\label{appendix:AGN}

Here we explore the possible bias introduced into the X-ray-based cross-spectra by AGN.  We test this by introducing an additional mask component which covers the regions with confirmed AGN sources. 

The second ROSAT all-sky survey point source catalogue\footnote{https://heasarc.gsfc.nasa.gov/W3Browse/rosat/rass2rxs.html} (2RXS, \citealt{Boller_etal_2016}) contains the most complete list of point-like sources observed with the PSPC in the full energy range of RASS. This catalogue also cross-matches the observed sources against a catalogue of known AGNs by \citet{Veron-Cetty_Veron_2010}.   Some $\sim 8000$ sources are confirmed as AGN to within 1 arcmin of the original source.  We use these sources to construct two new masks covering different number of pixels around the AGN.

Note that the 0.1-2.4 keV flux limit of the 2RXS RASS point source catalog is $\approx 10^{-13}$ ergs/s/cm$^2$ \citep{Boller_etal_2016}.  Taking the maximum redshift of our deeper SDSS photometric sample ($z=0.15$), this corresponds to a conservative soft X-ray luminosity limit of $\approx6.1\times10^{42}$ ergs/s.  For the redshift range $0.015$--$0.2$, \citet{Miyaji2001} find that $L_{X,*}$ (the characteristic AGN soft X-ray luminosity) in the same band is $\approx3.6^{+4.4}_{-2.0}\times10^{43}$ ergs/s (assuming $H_0=70$ km/s/Mpc).  \citet{Hasinger2005} combined various ROSAT surveys with deeper Chandra and XMM-Newton observations to derive a more precise (but consistent) constraint of $L_{X,*} = 2.82^{+3.07}_{-1.30}\times10^{43}$ ergs/s over the same redshift range.  Thus, the 2XRS catalog typically probes about a factor of 5 below the characteristic AGN soft X-ray luminosity and should therefore capture most of the X-ray AGN signal in this low redshift regime.  Using the luminosity function data in Table 3 of \citet{Hasinger2005}, we estimate that $\approx90\%$ of the X-ray AGN signal lies above the 2XRS point source limit of $6\times10^{42}$ ergs/s when integrating from $10^{42}$ ergs/s up (the lower limit of the Hasinger luminosity functions).  However, we cannot exclude a possible non-negligible contribution from X-ray AGN with luminosities fainter than $10^{42}$ ergs/s.

In terms of our masking approach, we note that the combined resolution of ROSAT with the PSPC camera is $\approx 1.8$ arcmin. This is approximately equal to the resolution of the \texttt{HEALPix} maps. It would be reasonable to assume that a point source can be approximated as 1.8 arcmin in this case and, thus, assigned one pixel - this is our first mask. A more aggressive masking technique is to include the 8 neighbouring pixels as well, masking each source with an area of $\approx5\times5$ arcmin$^2$. These masks are then combined with the total mask composed of SDSS footprint, the \textit{Planck} tSZ Milky Way 40\% and point-source masks, and the RASS X-ray zero exposure mask.  

\begin{figure}
        \includegraphics[width=\linewidth]{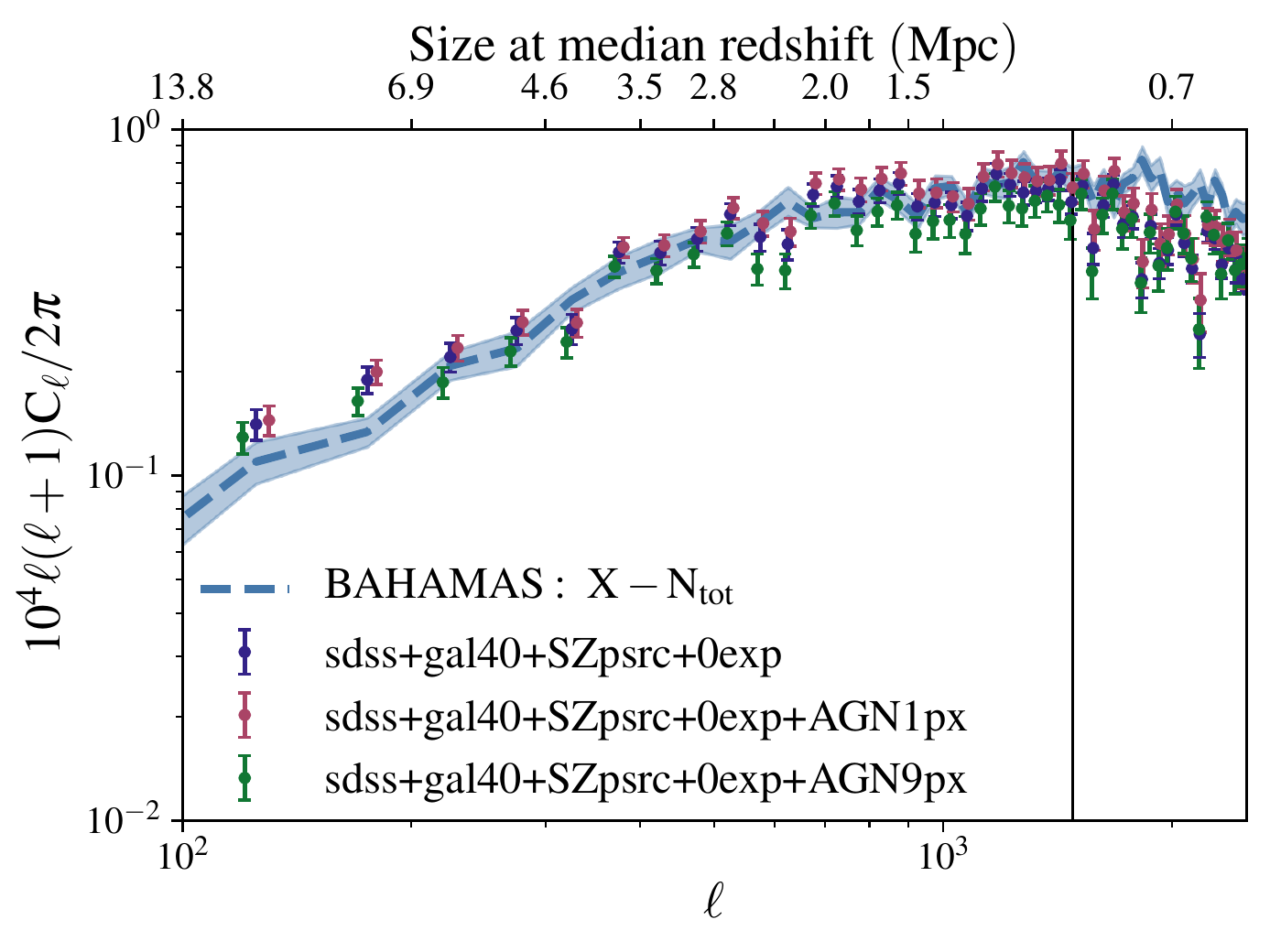}
    \caption{$\mathrm{X-\breve{N}_{tot}}$ cross-power spectra computed using different masks for AGN sources in the RASS x-ray map. Navy data points show the standard case from Figure~\ref{fig:panel_tsz_x} where AGN sources are not masked. Pink and navy data points represent 1 pixel and 9 pixel masks for AGN sources, respectively. Blue dashed line shows the equivalent power spectrum from BAHAMAS, which has been calibrated to reproduce hot gas fractions but does not contain AGN sources in X-ray emission. Data points have been artificially offset along the x-axis to make them more visible.  A small shift in the cross-power spectrum is visible when masking the AGN, particularly when the larger (9-pixel) mask is employed, suggesting AGN contribute at a sub-dominant level to the total (unmasked) cross-power spectrum.
    }
    \label{fig:AGN}
\end{figure}

Figure~\ref{fig:AGN} shows the resulting $\mathrm{X-\breve{N}_{tot}}$ cross-spectra for the AGN-unmasked case shown in Fig.~\ref{fig:panel_tsz_x} in navy, AGN-1px mask in pink, and AGN-9px mask in green. We observe a small difference in the measured cross-spectrum for both of the AGN masks relative to the unmasked case, implying that AGN do slightly contaminate our measured cross-spectrum.  However, the magnitude of the effect is not large enough to question the overall nature of the detection (i.e., it is dominated by hot diffuse gas) or to alter the main conclusions of our study.  Note that all three cross-spectra are still consistent over most angular scales with BAHAMAS, which has been calibrated to contain observed gas fractions inside groups and clusters and reproduces their X-ray scaling relations (see \citealt{McCarthy_etal_2017}). These two pieces of evidence lead us to conclude that AGN point sources are sub-dominant in the cross-correlations measured in this study, though with the caveat that the faint end of the AGN soft X-ray luminosity function is not well constrained below $10^{42}$ ergs/s.

\bsp	
\label{lastpage}
\end{document}